\documentclass{article}

\usepackage[utf8]{inputenc}
\usepackage[T1]{fontenc}
\usepackage[english]{babel}
\usepackage{amsmath, amsfonts}
\usepackage{eurosym}
\usepackage{graphicx}
\usepackage{amssymb}
\usepackage{color}
\usepackage[round]{natbib}
\usepackage{hyperref}
\usepackage{fullpage}
\newcommand{\N}{\mathbb{N}}
\newcommand{\R}{\mathbb{R}}

\newcommand{\X}{\mathcal{X}}
\newcommand{\textito}[1]{#1}%PARA CORRECCIONES

\newcommand{\textitr}[1]{\textcolor[rgb]{1.00,0.00,0.00}{\textbf{#1}}}%PARA CORRECCIONES
\newcommand{\textitb}[1]{\textcolor[rgb]{1.00,0.00,0.00}{\textbf{#1}}}%PARA CORRECCIONES
\newcommand{\textitm}[1]{\textcolor[rgb]{1.00,0.00,0.00}{\textbf{#1}}}%PARA CORRECCIONES
\newcommand{\textitor}[1]{\textcolor[rgb]{1.00,0.00,0.00}{\textbf{#1}}}%PARA CORRECCIONES
\newcommand{\textitf}[1]{\textcolor[rgb]{0.00,0.00,1.00}{\textbf{#1}}}%PARA CORRECCIONES
\renewcommand{\textit}[1]{\textcolor[rgb]{1.00,0.00,1.00}{\textbf{#1}}}%PARA CORRECCIONES

\renewcommand{\textitr}[1]{\textcolor[rgb]{0.00,0.00,0.00}{\textbf{#1}}}%PARA CORRECCIONES
\renewcommand{\textitb}[1]{\textcolor[rgb]{0.00,0.00,0.00}{\textbf{#1}}}%PARA CORRECCIONES
\renewcommand{\textitm}[1]{\textcolor[rgb]{0.00,0.00,0.00}{\textbf{#1}}}%PARA CORRECCIONES
\renewcommand{\textitor}[1]{\textcolor[rgb]{0.00,0.00,0.00}{\textbf{#1}}}%PARA CORRECCIONES
\renewcommand{\textitf}[1]{\textcolor[rgb]{0.00,0.00,0.00}{\textbf{#1}}}%PARA CORRECCIONES
\renewcommand{\textit}[1]{\textcolor[rgb]{0.00,0.00,0.00}{\textbf{#1}}}%PARA CORRECCIONES

\renewcommand{\textitr}[1]{\textcolor[rgb]{0.00,0.00,0.00}{{#1}}}%PARA CORRECCIONES
\renewcommand{\textitb}[1]{\textcolor[rgb]{0.00,0.00,0.00}{{#1}}}%PARA CORRECCIONES
\renewcommand{\textitm}[1]{\textcolor[rgb]{0.00,0.00,0.00}{{#1}}}%PARA CORRECCIONES
\renewcommand{\textitor}[1]{\textcolor[rgb]{0.00,0.00,0.00}{{#1}}}%PARA CORRECCIONES
\renewcommand{\textitf}[1]{\textcolor[rgb]{0.00,0.00,0.00}{{#1}}}%PARA CORRECCIONES
\renewcommand{\textit}[1]{\textcolor[rgb]{0.00,0.00,0.00}{{#1}}}%PARA CORRECCIONES

\newtheorem{remark}{Remark}
\begin{document}

\title{Be-CoDiS: A mathematical model to predict the risk of human diseases spread between countries. Validation and application to the 2014-15 Ebola Virus Disease epidemic.%\thanks{Grants or other notes
%about the article that should go on the front page should be
%placed here. General acknowledgments should be placed at the end of the article.}
}
%\subtitle{Be-CoDiS: A model for Ebola and other human diseases.}
\author{Benjamin Ivorra$^{a,*}$, Diène Ngom$^{b}$ and Ángel M. Ramos$^{a}$
% Martínez-López, Beatriz $^{b}$;\\ Fernández Carrión, Eduardo$^{a,c}$;  
%Sánchez-Vizcaíno, José-Manuel$^{c}$;\\
%; Iggidr, Abderrahman $^{e}$;
%Redondo, Juana L.$^{f}$;\\ Ortigosa, Pilar M.$^{f}$
\\~\\
$^{a}$ MOMAT research group, IMI-Institute and\\ 
Applied Mathematics Department.\\ 
Complutense University of Madrid.\\ 
Plaza de Ciencias, 3, 28040, Madrid, Spain.
\\~\\
%$^{b}$ UC Davis School of Veterinary Medecine.\\ 
%2415A TUPPER HALL Davis, CA 95616
%\\~\\
%$^{c}$ VISAVET Center and Animal Health Department.\\ 
%Veterinary School.
% Complutense University of Madrid.
%\\~\\
$^{b}$Département de Mathématiques\\
UFR des Sciences et Technologies\\
Université Assane Seck de Ziguinchor \& \\
 Laboratoire d'Analyse Numérique et d' Informatique,\\
					Université Gaston Berger de Saint Louis, Sénégal.
%\\~\\
%$^{e}$Université de Lorraine et IECL (CNRS)\\
% ISGMP, Ile du Saulcy, 57045 METZ \\~\\
%$^{f}$  Dept. de Inform\'atica, Universidad de Almer\'{\i}a,\\  
%ceiA3,  Ctra. Sacramento, s/n, 04120\\  
%La Ca\~nada de San Urbano, Almer\'{\i}a, Spain.
\\~\\
$^{*}$ Corresponding author: Tel: +34 91 394 44 15;\\
E-mail: ivorra@mat.ucm.es}
\date{Version 5.0 - Date of publication: \today}

\maketitle

\begin{abstract}
Ebola virus disease is a lethal human and primate disease that currently requires a particular attention from the international health authorities due to important outbreaks in some Western African countries and isolated cases in \textitf{the United Kingdom}, the USA and Spain. Regarding the emergency of this situation, there is a need of development of decision tools\textitf{, such as mathematical models,} to assist the authorities to focus their efforts in important factors to eradicate Ebola. In this work, we propose a novel deterministic spatial-temporal model, called Be-CoDiS (Between-Countries Disease Spread), to study the evolution of human diseases \textitf{within and between} countries. The main interesting characteristics of Be-CoDiS are the consideration of the movement of people between countries, \textitf{the control measure effects} and the use of time dependent coefficients adapted to each country. First, we focus on the mathematical formulation of each component of the model and explain how its parameters and inputs are obtained. Then, in order to validate our approach, we consider \textitf{two numerical experiments regarding the 2014-15 Ebola epidemic. The first one studies the ability of the model in predicting the EVD evolution between countries starting from the index cases in Guinea in December 2013. The second one consists \textit{of} forecasting the evolution of the epidemic by using some recent data.} The results obtained with Be-CoDiS are compared to real data and other models outputs found in the literature. Finally, a brief parameter sensitivity analysis is done.  \textitf{A free Matlab version of Be-CoDiS is available at:} \url{http://www.mat.ucm.es/momat/software.htm}
% \PACS{PACS code1 \and PACS code2 \and more}
% \subclass{MSC code1 \and MSC code2 \and more}
\end{abstract}

\section{Introduction}

Modeling and simulation are important decision tools that can be used to control or eradicate human and animal diseases \citep{Anderson,thieme}. Each disease presents its own characteristics and, thus, most of them \textitf{require} a well-adapted simulation model in order to tackle real situations \textitor{\citep{WN,Brauer}}.

In this work, we present a first version of a new \textitm{deterministic} spatial-temporal epidemiological model, called Be-CoDiS (\textitor{Between-Countries Disease Spread}), \textito{to simulate} the spread of human \textitm{diseases} in a considered area. This model is \textito{inspired from} a previous \textitm{one}, called Be-FAST (Between Farm Animal Spatial Transmission), which \textito{focuses} on the spread of animal diseases between and within farms. The major original ideas introduced by Be-FAST were the following: (i) the study of both within and between farms spread, (ii) the use of real database and (iii) dynamic coefficients calibrated in time according to farms characteristics (e.g., size, type of production, etc.). This model was deeply detailed in \citet{ivo1,ivo2} and validated on various cases as, for example: \textitor{Classical Swine \textitm{Fever}} in Spain and Bulgaria \citep{ivo3,ivo4}, and Foot-and-Mouth disease in Peru \citep{ivo5}. Be-CoDiS is based on the combination of a deterministic Individual-Based model (\textitf{with} the countries \textitm{playing the role of} individuals) \citep{DeAngelis}, simulating the between-country interactions (here, \textitr{movement of people between countries}) and disease spread, with a deterministic compartmental model \citep{Brauer} (a system of ordinary differential equations), simulating the within-country disease spread. We \textito{observe} that the coefficients of the model are calibrated dynamically according to the country indicators (e.g., economic situation, climatic conditions, etc.).  At the end of a simulation, Be-CoDiS returns outputs referring to outbreaks characteristics (for instance, \textitr{the basic reproductive ratio,} the epidemic magnitude, the risk of disease introduction or diffusion per country, the probability of having at least one infected per time unit, etc.). The  \textitf{main characteristics} of our approach \textitf{are} the \textitf{consideration of} \textitr{movement of people} between countries \textitf{and} control measure effects and \textitf{the use of} time dynamic coefficients fitted to each country.

This work has two main goals. The first one, is to give a full and detailed mathematical formulation of Be-CoDiS \textitr{and to explain how the model parameters are obtained} in order to provide a transparent and understandable model for users. The second one is to validate our model by applying it to the current case of the Ebola Virus Disease (EVD) \citep{ebo1,ebo2,ebo3}. 

EVD is a lethal human and primates \textito{disease} caused by the \textito{Ebola-virus} (family of the Filoviridae) that causes important clinical signs, such as haemorrhages, fever or muscle pain. The fatality percentage (i.e., the percentage of infected persons who do not survive the disease) is evaluated to be within 25\% and 90\%, due to hypovolemic shock and multisystem organ failure,  depending on the sanitary condition of the patient \textito{and the medical treatment}. This virus was first identified in Sudan and Zaire in 1976 (see \citet{ebo1}).  Various important outbreaks \textito{occurred} in 1995, 2003 and 2007 in the Democratic Republic of Congo (315, 143 and 103 persons infected by EVD, respectively), in 2000 and 2007 in Uganda (425 and 149 persons infected by EVD, respectively), see \citet{ebo2,ebocongo}. The \textitr{2014-15} outbreak started in December 2013 in Guinea and spread to Liberia and Sierra Leone. In March 2014, the international community was \textito{aware of} the gravity of the situation in those three countries. The situation on \textitf{April 24$^{\rm th}$, 2015} (the date used to run our numerical experiments) was a total of \textitf{26307} persons infected by EVD in Guinea, Liberia, Sierra Leone and Nigeria (see \citet{ebo3,ebo4}).  Moreover, \textitr{15} isolated cases were detected in Mali, Senegal, Spain, \textitr{the United Kingdom} and the USA. Furthermore, in Spain\textitr{, the United Kingdom} and the USA, the first contagions between people outside Africa were observed. \textito{The observed mean fatality percentage for this particular hazard decreased from \textitr{72.8\%} (on March, 2014) to \textitr{47.5\% }(on April, 2015), \textito{see \citet{Ebolaanimal}}}. From an epidemiological point of view, the EVD can be \textito{transmitted between} natural reservoirs (for instance, bats) and humans due to the contact with animal carcass \citep{Ebolaanimal}. The most common way of EVD human transmission is due to contacts with blood or bodily fluids from an infected person (including dead \textito{persons}).

%EVD is a human and primates virus disease that causes a high mortality rate (between 50\% and 90\%). Currently, several important outbreaks have been reported in Western Africa (Guinea, Liberia, Sierra Leone and Nigeria). Furthermore, seven isolated cases were detected in Mali, Senegal, the USA and Spain and the risk to have a EVD propagation outside Africa (its natural reservoir) is real.
 
Starting from this particular context, we study the behaviour of our model in predicting the possible spread of EVD worldwide. In order to validate our approach, we first consider a numerical experiment starting from the index cases in Guinea in December 2013 and check the ability of the model in spreading the \textitb{EVD} to other countries. Then, we perform a second experiment which aims to forecast the possible evolution of \textitb{EVD} by considering recent data.  \textitr{Moreover, we also simulate the possible evolution of \textitb{EVD} in the countries with the highest risks of \textitb{EVD} introduction due to the movement of people.} The results obtained by Be-CoDiS at the end of those experiments are compared to historical data and with other studies found in the literature \textito{\citet{ebo7,ebo5,ebocomp,ebo9,ebo3,ebo4,ebort} }regarding the same \textitr{2014-15} EVD outbreaks. The work in \citet{ebocomp} is based on a time spatial model with \textitm{stochastic} flux between areas (considering a database based on airport traffic instead of, as done here, \textitr{calibrated values of migratory flux}) but with a different point of view regarding control measures (at the beginning of the simulation, if control measures are applied the authors set the disease transmission rate to a value lower than in the case without control measures) and model coefficients (considered as constant in time). We point out that most of the parameters used by our model are calibrated for African countries and few data are available about the behaviour of EVD in other  countries. Thus, some empirical hypothesis are needed and a sensitivity analysis of Be-CoDiS regarding those parameter is done. Finally, we also highlight the current limitations of the model and a way to improve it in future works. 

This work is organized as follows. In \textito{Section \ref{math}, we describe the epidemiological behavior of EVD and give a detailed } presentation of our model. \textitr{In particular, we focus on its mathematical formulation and explain how its parameters are obtained.} %and analyze, in a simplified case, its behavior. 
\textitr{In Section \ref{num}, we focus on the validation of Be-CoDiS by simulating possible evolutions of the \textitb{EVD} spread worldwide and by comparing the obtained results with observed data and other studies.} Finally, we study the behavior of our model regarding changes in \textitr{the whole set of values} of its parameters.

\section{Mathematical formulation of the model \label{math}}
\textito{In Section \ref{ebola}} \textito{\textito{we} detail the epidemiological characteristics of EVD that are taken into account in our model. \textito{Then, in Sections \ref{gd}, \ref{WHT} and \ref{BHT}},} we describe in detail the Be-CoDiS model by presenting its general structure, \textito{i.e.,} the considered within and between countries disease spread sub-models. %and by studying its behavior regarding its stability points.
\textito{Finally, in Section \ref{cout}, we present the outputs used to analyse \textito{the results of the numerical} simulations presented in Section \ref{num}. } 
The main notations used in this work are summarized in Table \ref{tnota}.

\begin{remark}
Be-CoDiS is designed to be able to study the spread of any human disease worldwide. Here, some particular details of the model are related to the specific EVD case but it can \textito{be adapted} to other disease. For instance, the classification of compartments in the SEIHRDB model can be changed to study other cases. 
\end{remark}

\begin{table}[h!tb]
\begin{center}
\textitr{
\caption{Summary of the main notations used in this work to describe Be-CoDiS. A brief description (\textbf{Description}) and the range of the considered values (\textbf{Value}) are also reported. The reference are given in the corresponding \textito{section} of the text.}
\begin{tabular}{lllccccccc}
\hline
\textbf{Notation}&\textbf{Value} &\textbf{Description}\\
\hline
$T_{\max}$ &[0,+$\infty$)& Maximum number of simulation days (day)\\ 
$\Delta t$ & 1 & Time discretisation step size (day)\\ 
\textito{$\beta_{I}(i)$} & [0.1657,0.2671]& Disease contact rate of a person\\
&& in state $I$ in country $i$ (day$^{-1}$)\\
\textito{$\beta_{H}(i)$} & \textito{$0.04 \cdot \beta_{I}(i)$} & Disease contact rate of a person\\
&& in state $H$ in country $i$ \textitb{(day$^{-1}$)}\\
\textito{$\beta_{D}(i)$} & \textito{$\beta_{I}(i)$} & Disease contact rate of a person\\
&& in state $D$ in country $i$ (day$^{-1}$)\\
$\gamma_{\rm E}$ &  0.0877 & Transition rate of a person in state $E$ (day$^{-1}$)\\
$\gamma_{\rm I}(i,t)$ & [0.2,0.5]& Transition rate of a person in state $I$(day$^{-1}$)\\
&& in country $i$ at time $t$,\\ 
\textitb{$\gamma_{\rm HR}(i,t)$}& [0.14,0.2]& Transition rate of a person in state $H$\\
&& to state $R$ (day$^{-1}$) in country $i$ at time $t$,\\
\textitb{$\gamma_{\rm HD}(i,t)$} &  [0.13,0.24]&Transition rate of a person in state $H$\\
&& to state $D$  (day$^{-1}$) in country $i$ at time $t$, \\
\textito{$C_{o}$} & 12.9&\textito{the period of convalescence (day)}\\
$\mu_{m}(i)$ &[0.012,0.023]& Natural mortality rate in country $i$ (day$^{-1}$)\\
$\mu_{n}(i)$ &\textitm{[0.22,1.37]$\cdot10^{-4}$}& Natural natality rate in country $i$ (day$^{-1}$)\\
$\omega(i,t)$ &[0.25,0.728]& Disease fatality percentage in country $i$ at time $t$\\
\textito{$\kappa_i$} &[0.001,0.281]&  \textito{Efficiency of the control measures}\\ 
&&\textito{in country $i$ (day$^{-1}$)}\\
\textito{$\lambda(i)$}  &[0,+$\infty$)& \textito{First day of application of}\\ 
&&\textito{control measures in country $i$ (day)}\\ 
$m_I(i,t) / $  &[0,1]& Control measure efficiency (\%) in country $i$ at \\
$m_H(i,t)$         && time $t$  applied to persons in state $I$ or $H$\\
$m_{\rm tr}(i,j,t)$  &[0,1]& Control measure efficiency (\%) applied to persons\\
   && \textitr{moving} from country $i$ to country $j$ at time $t$\\
$\tau(i,t)$ && Daily \textitb{rate of the movement of people} \\
&& \textitr{from country $i$} to country $j$ (day$^{-1}$)\\
\textitf{$\delta$}&\textitf{ 0.53}&\textitf{Proportion of the $\omega(i,t)$ that can be reduced}\\
&&\textitf{due to the application of control measures}\\
$N_{\rm CO}$ &176& \textitb{Number of countries}\\
$NP(i,t)$ &\textitm{[2.5$\cdot10^{5}$,1.4$\cdot10^{9}$]}& Number of persons in country $i$ at time $t$\\
$S(i,t) / E(i,t) $ & \textitm{[2.5$\cdot10^{5}$,1.4$\cdot10^{9}$]} & Number of persons in state $S$, $E$, $I$, $H$, $R$, $D$, $B$\\
$I(i,t)  / H(i,t) $ && in country $i$ at time $t$\\
$R(i,t) / D(i,t)$&&\\
$B(i,t)$\\&&\\
\hline
\label{tnota}
\end{tabular}}
\end{center}
\end{table}

\subsection{\textito{Epidemiological} characteristics of Ebola Virus Disease \label{ebola}}

\textito{When} a person is not infected by EVD,  it is categorized in the Susceptible state (denoted by $S$). If a person is infected, then they passes successively through the following states (see \citet{Ebodur,ebo9,ebo2,ebort}):  

\begin{itemize}
\item Infected (denoted by $E$): The person is infected by EVD but they cannot infect other people and has no visible clinical signs (i.e., fever, hemorrages, etc.). The mean duration of a person in this state is 11.4 days (range of [2,21] days) and is called incubation period. Then, the person passes to \textito{the} infectious state. %C4 has

\item Infectious (denoted by $I$): The person can infect other people and start developing clinical signs. \textitr{The mean duration of a person in this state is called infectious period. \textitb{In} September\textitb{,} 2014, the mean infectious period \textitm{was} \textito{5} days (range of [0,10] days\textitm{, according to} \cite{ebort}). However, due to the control measures applied to fight the \textitb{EVD} epidemic presented below, \textitb{in} March\textitb{,} 2015, the mean infectious period  is estimated to be between [2.6,3] days for affected African countries \textitb{(see, \cite{ebo4})}. Furthermore it \textitb{took} only 2 days to hospitalize an infected patient in the United Kingdom.} After this period, \textitb{infectious persons are} taken in charge by sanitary authorities and we classify \textitb{them} as Hospitalized.

\item Hospitalized (denoted by $H$): The person is hospitalized and can still infect other people, but with a lower probability. The mean duration in this sate is 4.5 days (see \citet{Ebodur}). Actually, it has been observed in practice that those patients can still infect other people with a probability 25 times lower than infectious people (see \citet{ebo4}). On the one hand, after a mean duration of 4.2 days (range [1,11] days), \textito{a percentage of the hospitalized persons, between 25\% and 72.8\% (depending on the sanitary services of the country),} die due to the EVD clinical signs and pass to the \textito{Dead state}. On the other hand, \textito{after a mean duration of 5 days,} the persons who have survived to EVD pass to the Recovered state. \textito{We \textito{point out} that \textito{state $H$} does not include hospitalized persons which cannot infect other people \textito{any more}. This last category of persons is included in the recovered state explained below. \textito{The mean of the total number of \textito{days that a EVD patient \textito{stays} in }hospital (from hospitalization to hospital discharge) is estimated to be \textito{11.8 days (range [7.7,17.9] days)}.}}

\item Dead (denoted by $D$): The person has not survived to the \textito{EVD}. %The mean probability of dying due to those clinical signs is 70.8\% (range of [68,73]\%). As the mean duration between the clinical signs onset and the death is 7.5 days (range [1,15] days), we assume that the death process occurs at the Hospitalized state. 
It has been observed in previous Ebola epidemics, that cadavers of infected persons can infect other people until they are buried. The probability to be infected by this kind of contact is the same as \textito{that of being infected}  by contact with infectious persons. \textito{Indeed, the daily number of \textito{contacts} of a \textito{cadaver with persons} is lower than \textito{that of an} infectious \textito{person} but, the \textito{risk of EVD transmission} due to \textito{contacts with} cadavers is larger than the risk due to \textito{contacts with} infectious persons (see \citet{Ebodur}).} After a mean period of 2 days \textito{(this could be different for some countries)} the body is buried.

\item Buried (denoted by $B$): The person is dead \textito{because of EVD}. Its cadaver is buried and \textitm{they can} no longer \textitm{infect other people}. 

\item Recovered (denoted by $R$): The person has survived \textito{\textito{the} EVD} and is no longer infectious. \textito{They develop} a natural immunity to EVD. Since it has never been observed a person who has recovered from Ebola and contracted the disease again during the period of time of the same epidemic, it is assumed that they cannot be infected \textito{again} by Ebola. 

\end{itemize}

%A person in the state $E$, $I$ or $H$ is called a contaminated person. Moreover, a person in the state $I$ or $H$ is also called a spreading person.

Once an EVD infected person is hospitalized, the authorities apply various control measures in order to control the EVD spread (\textito{see \citet{ebo5,controlchav,ebo3,ebo4}}):
\begin{itemize}
\item Isolation: Infected people are isolated from contact with other people. Only sanitary professionals are in contact with them. However, contamination of those professionals also occur (see \citet{ebo5}). \textito{Isolated persons receive an adequate medical treatment that reduces the EVD fatality rate.}

\item Quarantine: Movement of people in the area of origin of an infected person is restricted an controlled (e.g., quick sanitary check-points at the airports) to avoid that possible infected persons spread the disease.

\item Tracing: The objective of tracing is to identify potential infectious contacts which may have infected a person or spread EVD to other people. 

\item Increase of sanitary \textitr{resources: The number of operational beds and sanitary personal available to detect and treat affected persons is increased, producing a decrease in the infectious period.} This greatly the time needed to  The funerals of infected cadavers are controlled by sanitary personal in order to reduce the contacts between the dead bodies and susceptible persons. 
\end{itemize}

%We note that once a person is detected\citep{ebo4}.

\begin{remark}
\textito{Data given above were} calibrated for the cases of African countries, the natural reservoir of EVD. However, due to the spread of this disease out of Africa, new studies should be performed to analyze the behavior of EVD in other sanitary, population and climatic conditions. Currently, very few studies are available. One of them is about the survival of the Ebola virus (EV) according to changes in temperature (see \citet{ebo8}). It has been found that the lower is the temperature the greater is the survival period of the EV) outside the host. Thus, in this work some empirical hypothesis, which seems to be reasonable, have been done. We have assumed that the transmission parameter of EV decreases when the temperature or the sanitary expenses of a country increase and increases when the people density of a country increases. 
\end{remark}

\subsection{General description \label{gd}}

The Be-CoDiS model is used to evaluate the spread of \textito{a human} disease within and between countries during a fixed time interval. 

At the beginning of the simulation, the model parameters are set by the user (for instance, the values considered \textitm{for EVD}  are described in Section \ref{parameter}). At the initial time ($t=0$),  only susceptible people live in the countries that are free of \textito{the disease}, whereas the number of persons \textito{in states $S$, $E$, $I$, $H$, $R$, $D$ and $B$} of the infected countries are set to their \textitm{corresponding values.} Then, during the time interval $[0,T_{\max}]$, with $T_{\max} \in {\rm I}\! {\rm N}$ \textitm{being} the maximum number of simulation days, the within-country and between-country daily spread procedures (described in Section \ref{WHT}) are applied. If at the end of a simulation day $t$ all the people in \textito{all} the considered countries are in the susceptible state, the simulation is stopped. Else, the simulation is stopped when $t= T_{\max}$. Furthermore, the control measures are also implemented and they can be activated or deactivated, when starting the model, in order to quantify their effectiveness to reduce the magnitude and duration of a EVD epidemic.

\begin{figure}[!htb]
\begin{center}
\includegraphics[width=11.89cm]{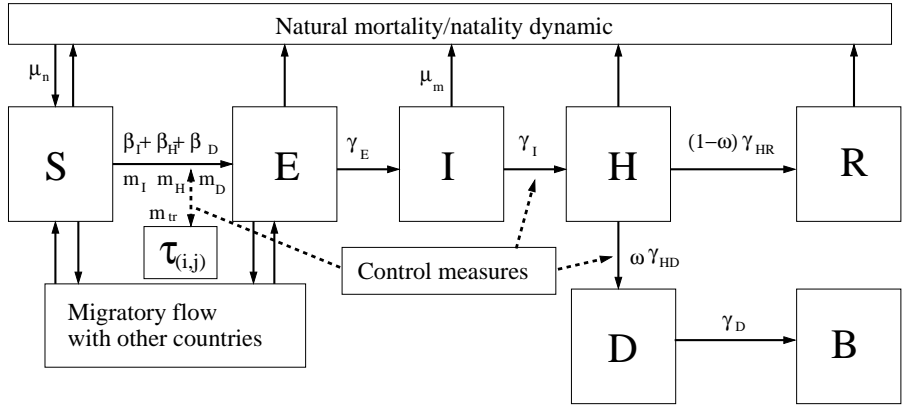}
\caption{Diagram summarizing the Be-CoDiS model.\label{diag}}
\end{center}
\end{figure}

A diagram summarizing the main structure of our model is presented in Figure \ref{diag}.

\textitr{\begin{remark}
The choice of using a deterministic model instead of a stochastic one is done as a first approach, \textitb{since such kind} of models presents some advantages, such as: a low computational complexity allowing a better calibration of the model parameters or the possibility \textitb{of using} the theory of ordinary differential equations for well analyzing and interpreting the model. Furthermore, according to \citet{diekmann}, deterministic models should be the first tool \textitb{to be} used when modelling a new problem with few data (here, few data of the spread of Ebola are available for non-African countries). The authors \textitb{of that work} also note that the stochastic models are not \textitb{suitable} when it is difficult or impossible to determine the distribution probability, are difficult to analyze and require more data for the calibration of the model.   
\end{remark}}

\subsection{\textitb{Within-Country} disease spread \label{WHT}}

The dynamic disease spread within a particular contaminated country $i$ is modeled by using a deterministic compartmental model (see\textitf{, for instance, \cite{Brauer}}). 

We assume that the people in a country are characterized to be in one of those states, described in Section \ref{ebola}: Susceptible ($S$), Infected ($E$), Infectious ($I$), Hospitalized ($H$),  Recovered  ($R$)\textito{, Dead ($D$) or Buried ($B$)}. For the sake of simplicity we assume that, at each time, the population inside a country is homogeneously distributed (this can be improved by dividing some countries into a set of smaller regions with similar characteristics). Thus, the spatial distribution of the epidemic inside a country can be omitted. We \textitm{also} assume that new births are susceptible persons. %such that the country population is constant (i.e., each death gives a birth in the susceptible state). 
In Section \ref{WHT}, we do not consider interaction between countries.

\textito{Under those assumptions,} the evolution of $S(i,t)$, $E(i,t)$, $I(i,t)$, $H(i,t)$, $R(i,t)$, \textito{$D(i,t)$ and $B(i,t)$,} denoting the number of susceptible, infected, infectious, hospitalized, recovered,\textito{ dead and buried }persons in country $i$ at time $t$, respectively, is modeled by \textito{the following system of ordinary differential equations}
\begin{eqnarray} \label{SEIHR}
\begin{array}{rcl}
\dfrac{{\rm d} S(i,t)}{{\rm d}t}&=& - \dfrac{S(i,t) \bigg( m_{I}(i,t)\beta_I(i) I(i,t)+m_{H}(i,t)\beta_H(i) H(i,t)  \bigg) }{NP(i,t)}\\~~\\
&& - \dfrac{S(i,t) \bigg( m_{D}(i,t)\beta_D(i) D(i,t) \bigg) }{NP(i,t)} - \mu_m(i) S (i,t) \\~~\\
&& + \mu_n(i) \bigg(S(i,t)+ E(i,t) +I(i,t) + H(i,t) +R(i,t)\bigg)   , \\~~\\ 
\dfrac{{\rm d} E(i,t)}{{\rm d}t}&=&  \dfrac{S(i,t)\bigg(m_{I}(i,t)\beta_I(i) I(i,t)+ m_{H}(i,t)\beta_H(i) H(i,t) \bigg) }{NP(i,t)}\\~~\\ 
&& + \dfrac{S(i,t)\bigg(m_{D}(i,t)\beta_D(i) D(i,t)\bigg) }{NP(i,t)} - \mu_m(i)  E(i,t) \\~~\\ 
&& - \gamma_E  \X_{\epsilon_{\rm fit}}(E(i,t)) ,\\~~\\ 
\dfrac{{\rm d} I(i,t)}{{\rm d}t}&=&  \gamma_E \X_{\epsilon_{\rm fit}}(E(i,t)) - (\mu_m(i) + \gamma_I)(i,t) I(i,t),\\~~\\ 
\dfrac{{\rm d} H(i,t)}{{\rm d}t}&=&  \gamma_I(i,t) I(i,t) \\~~\\
&&- \bigg(\mu_m(i) + (1-\omega(i,t) ) \gamma_{HR}(i,t) + \omega(i,t) \gamma_{HD}(i,t)  \bigg) H(i,t),\\~~\\
\dfrac{{\rm d} R(i,t)}{{\rm d}t}&=&  \big(1-\omega(i,t) \big) \gamma_{HR}(i,t) H(i,t) - \mu_m(i) R(i,t), 
\quad \quad \quad \quad \quad \quad \quad \quad \quad  \\~~\\  
\dfrac{{\rm d} D(i,t)}{{\rm d}t}&=&  \omega(i,t) \gamma_{HD}(i,t) H(i,t)  - \gamma_D D(i,t),\\~~\\  
\dfrac{{\rm d} B(i,t)}{{\rm d}t}&=&  \gamma_D D(i,t) ,
\end{array}
\end{eqnarray}
where:
\begin{itemize}
\item $i\in \{1,\ldots, N_{CO}\}$, 
\item $N_{\rm CO} \in \N$ is the number of countries,
\item \textito{$NP(i,t)= S(i,t)+E(i,t)+I(i,t)+H(i,t)+R(i,t)+D(i,t)+B(i,t)$} is the number of persons \textito{(alive and also died or \textito{buried} because of EVD) }in country $i$ at time $t$,
\item $\mu_n(i) \in [0,1]$ is the natality rate (day$^{-1}$) in country $i$: the number of births per day and per capita, 
\item $\mu_m(i) \in [0,1]$ is the mortality rate (day$^{-1}$) in country $i$: the number of deaths per day and per capita (or, equivalently, the inverse of the mean life expectancy (day) of a person),
\item \textito{$\X_{\epsilon_{\rm fit}}(x)=x$ if $x\geq \epsilon_{\rm fit}$, \textitm{$\X_{\epsilon_{\rm fit}}(x)=2  x-\epsilon_{\rm fit}$ if $(\epsilon_{\rm fit}/2) \leq  x\leq \epsilon_{\rm fit}$}, and 0 elsewhere, with \textitm{$\epsilon_{\rm fit} \geq 0$} (small tolerance parameter).} \textitr{This function is a filter used to avoid artificial spread of the epidemic due to negligible values of $x$,}

\item $\omega(i,t) \in [0,1]$ is the disease fatality percentage in country $i$ at time $t$: the percentage of persons who do not survive the disease,
\item $\beta_I(i) \in \R^+$ is the disease \textito{effective} contact rate (day$^{-1}$) \textito{of a person in state $I$} \textito{in country $i$}: the mean number of effective contacts (i.e., \textito{contacts sufficient} to transmit the disease) of a person \textito{in state $I$} per day before applying control measures,
\item $\beta_H(i) \in \R^+$ is the disease \textito{effective} contact rate (day$^{-1}$) \textito{of a person in state $H$} in \textito{country $i$},
\item $\beta_D(i) \in \R^+$ is the disease \textito{effective} contact rate (day$^{-1}$) \textito{of a person in state $D$} in \textito{country $i$},
\item \textitm{$\gamma_E(i,t)$}, \textitr{$\gamma_I(i,t)$, $\gamma_{HR}(i,t)$, $\gamma_{HD}(i,t)$}, \textitm{$\gamma_D(i,t) \in (0,+\infty)$} denote the transition rate (day$^{-1}$) from a person \textito{in state $E$, $I$, $H$, $H$ or $D$ to state $I$, $H$, $R$, $D$ or $B$}, respectively: the number of persons per day and per capita passing from one state to the other (or, equivalently, the inverse of the mean duration of one of those persons in state $E$, $I$, $H$, \textito{$H$,} or $D$, respectively). \textitr{We note that $\gamma_I(i,t)$, $\gamma_{HR}(i,t)$ and $\gamma_{HD}(i,t)$ are time and country \textitb{dependent, \textitm{since,} due} to the applied control \textitb{measures} in country $i$\textitm{,} their value \textitb{could} vary in time,}
\item $m_{I}(i,t)$,  $m_{H}(i,t)$, $m_{D}(i,t)\in[0,1]$ (\%) are functions representing the efficiency of the control measures applied to non-hospitalized persons, hospitalized persons and infected cadavers respectively, in country $i$ at time $t$ to eradicate the outbreaks. Focusing on the application of the control measures, which in the EVD consists in isolating persons/areas of risks and in improving sanitary conditions of funerals, we multiply the disease contact rates (i.e., $\beta_I(i)$, $\beta_H(i)$ and $\beta_D(i)$) by decreasing functions simulating the reduction of the number of effective contacts as the control measures efficiency is improved. Here, we have considered the functions (see \citet{cmapp}):
\begin{equation} \label{wcm}
m_{I}(i,t)=m_{H}(i,t)=m_{D}(i,t)=\exp \bigg(-\kappa_i \max(t-\lambda(i),0) \bigg),
\end{equation}
where \textito{$\kappa_i \in [0,+\infty)$} (day$^{-1}$) simulates the efficiency of the control measures (greater value implies lower value of disease contact rates) and $\lambda(i) \in \R \cup \{ + \infty\}$ (day) denotes the first day of application of those control measures. 
\end{itemize}

System \eqref{SEIHR} is completed with initial data $S(i,0)$, $E(i,0)$, $I(i,0)$, $H(i,0)$, $R(i,0)$\textito{, $D(i,0)$ and $B(i,0)$} given in $\R$, for $i$=1,.., $N_{CO}$.

We \textito{observe} that all parameters of \textito{system} \eqref{SEIHR} should be adapted to the considered disease and countries. Generally, they are calibrated considering real data as explained in Section \ref{num} for the EVD case. 

\textitr{\begin{remark}
Numerical experiments presented in Section \ref{res} \textitb{seem} to show that, despite its apparent simplicity, using \textitb{system} \eqref{SEIHR} to simulate the \textitb{within-country} disease spread gives reasonable results. However, a next step could be to model the \textitb{wihtin-country} spread by considering, for instance, an individual based model that simulates the interactions between communities (e.g., cities, villages, etc.), as done in the case of herds and livestock diseases in \citet{ivo5}. However, to do so, we first need to collect data about those communities (i.e., size, location, type, etc.) and their interacting network (i.e., movement of people, etc.). Obtaining those data at the worldwide level seems to be quite difficult and may require a large effort of data-mining. Furthermore, the computational time needed to solve such a model \textitb{would be much higher than that of} the current version of Be-CoDiS and, thus, the estimation of the model parameters \textitb{would} be complicated. However, some of the interests in developing this kind of complex model are, for instance, their ability to estimate the local efficiency of the control measures and to assess the geographical allocation of the control measures resources in order to contain efficiently the epidemic spread.   
\end{remark}}

\subsection{Between-Countries disease spread \label{BHT}}

The disease spread between countries is modeled by using a spatial deterministic Individual-Based model (see \citet{DeAngelis}). 
\textito{Countries} are classified in one of the following states: free of disease (\textito{$F$}) or with outbreaks (\textito{$O$}).
\textito{We assume that }at time $t$ country $i$ is \textito{in state $O$} if $I(i,t)+H(i,t) \geq 1$ (i.e., there exist at least one infected person in this country), else it is \textito{in state $F$}.

In this work we consider that the flow of people between countries $i$ and $j$ at time $t$ (i.e., \textitm{persons} traveling per day from $i$ to $j$ at time $t$), is the only way to introduce the disease from country $i$, in state \textito{$O$}, to country $j$. To do so, we consider the matrix $(\tau(i,j))_{i,j=1}^{N_{\rm CO}}$, where $\tau(i,j) \in[0,1]$ is the rate of transfer (day$^{-1}$) of persons  from country $i$ to country $j$, which is expressed in \% of population in $i$ per unit of time. Furthermore, we assume that only persons in the $S$ and $E$ sates can travel, as other categories are not in condition to perform trips due to the importance of clinical signs or to quarantine. Moreover, due to control measures in both $i$ and $j$ countries, we assume that those rates can vary in time and are multiplied by \textito{a function denoted by $m_{\rm tr}(i,j,t)$.} 

Thus, we consider the following modified version of \textito{system} \eqref{SEIHR}:

\begin{eqnarray} \label{SEIHRc}
\begin{array}{rcl}
\dfrac{{\rm d} S(i,t)}{{\rm d}t}&=& - \dfrac{S(i,t)\bigg( m_{I}(i,t)\beta_I(i) I(i,t)+m_{H}(i,t)\beta_H(i) H(i,t)  \bigg) }{NP(i,t)} \quad \quad\\~~\\
&& - \dfrac{S(i,t)\bigg( m_{D}(i,t)\beta_D(i) D(i,t) \bigg) }{NP(i,t)} - \mu_m(i) S(i,t)\\~~\\
&&  + \mu_n(i)\bigg( S(i,t) + E(i,t) +I(i,t) + H(i,t) +R(i,t)\bigg) \\~~\\ 
&&  + \sum_{i \neq j} m_{\rm tr}(j,i,t) \tau(j,i)  S(j,t)   - \sum_{i \neq j} m_{\rm tr}(i,j,t) \tau(i,j) S(i,t),\\~~\\ 
\dfrac{{\rm d} E(i,t)}{{\rm d}t}&=&  \dfrac{S(i,t)\bigg(m_{I}(i,t)\beta_I(i) I(i,t)+ m_{H}(i,t)\beta_H(i) H(i,t) \bigg) }{NP(i,t)}\\~~\\
&& + \dfrac{S(i,t)\bigg(m_{D}(i,t)\beta_D(i) D(i,t) \bigg) }{NP(i,t)} - \mu_m(i)  E(i,t)\\~~\\
&& + \sum_{i \neq j} m_{\rm tr}(j,i,t) \tau(j,i)  \X_{\epsilon_{\rm fit}}(E(j,t)) \\~~\\ 
&&- \sum_{i \neq j} m_{\rm tr}(i,j,t) \tau(i,j) \X_{\epsilon_{\rm fit}}(E(i,t)) - \gamma_E \X_{\epsilon_{\rm fit}}(E(i,t)),\\~~\\ 
\dfrac{{\rm d} I(i,t)}{{\rm d}t}&=&  \gamma_E \X_{\epsilon_{\rm fit}}(E(i,t)) - (\mu_m(i) + \gamma_I(i,t)) I(i,t),\\~~\\ 
\end{array}
\end{eqnarray}
\begin{eqnarray*} 
\begin{array}{rcl}
\dfrac{{\rm d} H(i,t)}{{\rm d}t}&=&  \gamma_I(i,t) I(i,t) \\~~\\
&&- \bigg(\mu_m(i) + (1-\omega(i,t) ) \gamma_{HR}(i,t) + \omega(i,t) \gamma_{HD}(i,t)  \bigg)H(i,t),\\~~\\
\dfrac{{\rm d} R(i,t)}{{\rm d}t}&=&  (1-\omega(i,t)) \gamma_{HR}(i,t) H(i,t) - \mu_m(i) R(i,t),\\~~\\  
\dfrac{{\rm d} D(i,t)}{{\rm d}t}&=&  \omega(i,t) \gamma_{HD}(i,t) H(i,t)  - \gamma_D D(i,t),\\~~\\  
\dfrac{{\rm d} B(i,t)}{{\rm d}t}&=&  \gamma_D D(i,t).
\end{array}
\end{eqnarray*}
System \eqref{SEIHRc} is completed with initial data $S(i,0)$, $E(i,0)$, $I(i,0)$, $H(i,0)$, $R(i,0)$, $D(i,0)$ and $B(i,0)$ given in \textito{$[0,+\infty)$}; for $i$=1,.., $N_{\rm CO}$. 

\textito{This full model \eqref{SEIHRc}, which is summarized in Figure \ref{diag}, is called Be-CoDiS.}

\begin{remark}
\textito{The explicit solution of \eqref{SEIHRc} and the corresponding initial values are not (in general) available. In order to get an approximation of the solution one can use a suitable numerical solver. Here, we use} the explicit Euler scheme with a time step of 1 day.
%\textito{Regarding the computational time, for instance, one simulation using this implementation, in Matlab 2014a and running on a laptop with a quadcore i7-3740QM 2.7GHz CPUs with 16Gb of ram,  with $T_{\max}$=365 days and the parameters given in Section \ref{valid} requires around 25 seconds.}
\end{remark}

\subsection{\textito{Outputs of the model} \label{cout}}

\textito{Here, we present the outputs used to analyse the results of the simulations performed in Section \ref{num}.
}
\textito{In particular, considering \textito{a time} interval $[0,T]$, for each country $i$ we  compute the following values:
}

\begin{itemize}

\item $\hbox{cumul$_{\rm cases}$}(i,t)$\textito{: the} cumulative number of EVD cases in country $i$ at day $t$, \textito{which can be computed} as:
\textito{$$
\hbox{cumul$_{\rm cases}$}(i,t)= \hbox{cumul$_{\rm cases}$}(i,0)+\int_0 ^t \gamma_I \cdot I(i,t) {\rm d}t.
%\approx \hbox{cumul$_{\rm cases}$}(i,0)+\sum_{k=1} ^t \gamma_I \cdot I(i,k),
$$}
%where $\hbox{cumul$_{\rm cases}$}(i,0)\in\R$ is given in \citep{ebo4}.

\item $\hbox{cumul$_{\rm deaths}$}(i,t)$: cumulative number of deaths (due to EVD) in country $i$ at day $t$, \textito{which can be computed as:}
\textito{$$
\hbox{cumul$_{\rm deaths}$}(i,t)= \hbox{cumul$_{\rm deaths}$}(i,0)+ \omega(i,t) \int_0 ^t  \gamma_{HD} \cdot H(i,t) {\rm d}t. 
%\approx \hbox{cumul$_{\rm deaths}$}(i,0)+ \omega(i) \sum_{k=1} ^t \gamma_{HD} \cdot H(i,k),
$$}
%where $\hbox{cumul$_{\rm deaths}$}(i,0)\in\R$ is given in \citep{ebo4}.

\item \textitr{R$_0$ and CR$_0(i)$: the basic reproductive ratio of the model and country $i$, respectively. It is defined as the number of cases one infected person generates on average over the course of its infectious period, in an otherwise uninfected population. \textitb{In} our model most of the parameters are time and country dependent, thus in order to approximate a value of R$_0$ we propose the following methodology. First, considering the approach proposed in \citet{ebocomp,Ebodur}, we compute $\overline{\rm CR_0}(i,t)$, an estimation of the the R$_0$ value obtained when considering \textitb{system }\eqref{SEIHR} with the parameters value of country $i$ at time $t$, given by
\begin{eqnarray*} 
\begin{array}{c}
\overline{{\rm CR}_0}(i,t)=\dfrac{m_{I}(i,t)\beta_I(i)}{\mu_m(i)+\gamma_I(i,t)}+\dfrac{\omega(i,t) m_{I}(i,t)\beta_D(i) }{\gamma_D(i,t)}\\~\\
+\dfrac{m_{I}(i,t)\beta_H(i) \gamma_I(i,t)}{\big(\mu_m(i)+\gamma_I(i,t)\big) \cdot \big(\mu_m(i) + (1-\omega(i,t) ) \gamma_{HR}(i,t) + \omega(i,t) \gamma_{HD}(i,t)  \big)}.
\end{array}
\end{eqnarray*} 
Then, according to the number of infected persons in each country at \textitb{time $t$,} we compute the mean basic reproductive ratio of the model at time $t$, denoted by $\bar{R0}(t)$, as
\begin{equation*}
\overline{\rm R_0}(t)={\rm mean}_{i=1,..,N_p} \bigg(\dfrac{{\rm CR}_0(i,t) \cdot \Big( E(i,t)+I(i,t)+H(i,t)+D(i,t) \Big)}{\sum_{j=1}^{N_p} E(i,t)+I(j,t)+H(j,t)+D(j,t) }\bigg).
\end{equation*}
Finally, we compute
\begin{equation}
{\rm R}_0={\rm mean}_{t\in[0,T_{\max}]} \overline{R_0}(t) \hbox{ and } {\rm CR}_0(i)={\rm mean}_{t\in[0,T_{\max}]} \overline{\rm CR_0}(i,t).
\end{equation}}

\item RS$(i)$:  the initial \textito{risk of country $i$} to spread EVD to other countries, given by:
\textitb{$$\hbox{RS}(i)= \sum_{i \neq j} \tau(i,j) m_{\rm tr}(i,j,0) E(i,0).$$}
RS (persons$\cdot$day$^{-1}$)  \textito{is} the daily amount of infected persons who \textito{leave  country} $j$ at time $t=0$. 

\item TRS$(i)$: the total risk of country $i$ to spread EVD to other countries, considering the time interval $[0,T]$,  computed as:
\textitb{$$ 
\hbox{TRS}(i)= \sum_{t=0}^{T} \sum_{i \neq j} \tau(i,j) m_{\rm tr}(i,j,t) E(i,t).
$$}
TRS (persons) \textito{is} the number of infected persons send to other countries during the considered time interval.

\item RI$(i)$:  the initial \textito{risk of EVD introduction into country $i$} from other countries, given by:
\textitb{$$\hbox{RI}(i)= \sum_{j \neq i} \tau(j,i) m_{\rm tr}(j,i,0) E(j,0).$$}
RI (persons$\cdot$day$^{-1}$) \textito{is} the daily amount of infected persons \textito{entering} country $i$ at time $t=0$.

\item TRI($i)$: the total risk of EVD introduction \textito{into country $i$} from other countries considering the time interval $[0,T]$, computed as:
\textitb{$$
\hbox{TRI}(i)= \sum_{t=0}^{T} \sum_{i \neq j} \tau(j,i) m_{\rm tr}(j,i,t) E(j,t).
$$}
TRI (persons) \textito{is} the number of EVD infected persons received from other countries during the time interval $[0,T]$.

\item MNH($i$): the maximum number of hospitalized persons at the same time at country $i$ during the time interval $[0,T]$. \textito{It is computed as:} 
$${\rm MNH}(i)=\max_{t=0...T}  \Big\{H(i,t) + R(i,t) - R(i,t-C_{o}) \Big\}.$$
We remind \textito{(see Section \ref{ebola}) that} $C_{o}$ is \textitr{the period of convalescence (i.e., the time \textitm{a} person is still hospitalized \textitm{after surviving EVD}).}
This number can help to estimate and plan the number of clinical beds needed to treat all the EVD cases. 

\item Emf$(i)$: \textito{the percentage of the population leaving the country each day (day$^{-1}$), which can be computed as:}
$$\hbox{Emf}(i)=\sum_{i \neq j}\tau(i,j).$$
\end{itemize}

\textitm{\begin{remark}
We note that $\hbox{cumul$_{\rm cases}$}(i,t)$ and $\hbox{cumul$_{\rm deaths}$}(i,t)$  are \textitf{some} of the main indicators reported in \cite{ebo4}, since March 2014, in order to give an estimation of the magnitude of the epidemic.
\end{remark}}

\section{Application to the \textitr{2014-15} Ebola case \label{num}}

We are now interested in validating our approach by considering the Ebola epidemic currently  occurring worldwide. The advantage of this case is that some real and simulated data are now available and, thus, we are able to compare our model outputs with the information available in the following literature \citet{ebo5,ebo6,ebo3}. 

To \textito{this aim}, \textito{we first explain} in Section \ref{parameter} how to estimate the model parameters for the EVD. \textito{Then,} we present the results and discuss them in Section \ref{res}. Finally, in Section \ref{san}, we carry out a brief sensitivity analysis regarding the parameters values.

\subsection{Be-CoDiS parameters estimation for EVD \label{parameter}} 

\textito{Some of the} parameters used in the simulations presented in Section \ref{res}, have been found in the literature \citet{ebo5,ebo6,ebocongo,ebocomp} and in the daily reports on the Ebola evolution available online (see \citet{ebo7,ebo4}). Despite the effort to use the maximum amount of robust parameters as possible, due to lack of information of the behavior of Ebola out of Africa, some of them have been estimated using empirical assumptions. This part should be clearly improved as soon as missing information is available.

We now detail each kind of parameter by its category.

\subsubsection{Country indicators \label{cind}}

\textito{We use the following data} regarding country $i$:
\begin{itemize}
\item $\hbox{TMP}_i(t) \in \R$: Mean temperature (ºC) at day $t$.
\item $\hbox{DEN}_i\in \R^+$: \textito{Population density (persons/km$^2$).}
\item $NP(i,0) \in \N$: Total number of persons \textito{alive and also died \textito{or buried} because of EVD \textitm{at time $t=0$ (see \cite{ebo4})}}.
\item $\hbox{GNI}_i \in \R^+$: Gross National Income per year per capita (US$\$$.person$^{-1}$.ye-ar$^{-1}$). We remind that the Gross National Income is an indicator of the economy \textito{of the} country: the total domestic and foreign output claimed by residents of a country.
\item $\hbox{SAN}_i \in \R^+$: The mean Health expenditure per year per capita (US$\$$.per-son$^{-1}$.year$^{-1}$). This is an economical indicator of the sanitary system of a country \textito{given by the amount of money inverted by public and private institutions (national or international) in the sanitary system of the country.}
\item $\hbox{MLE}_i \in \R^+$: The mean life expectancy (days).
\item $\mu_n(i) \in [0,1]$: See Section \ref{WHT}.
\end{itemize}
All those data have been freely obtained for year 2013 from the following World Data Bank website: \url{http://data.worldbank.org}. 
%We note that $\hbox{MLE}_i$, $\hbox{SAN}_i$, $\hbox{GNI}_i$, $\hbox{DEN}_i$ and $\hbox{DEN}_i$ are normalized respecting to their maximum value in order to be used in the formula presented below.
%A graphical representation of those data is presented in Figure \ref{?}.

\subsubsection{Initial conditions \label{estim}}

We have considered the initial conditions for our \textito{system} \eqref{SEIHRc} corresponding to the state of the EVD epidemic at \textito{several dates} reported in \citet{ebo4}. 

\textitr{In \citet{ebo4}, the cumulative} numbers of reported cases (i.e., \textito{persons who have ever been in} state $H$, as stated in \citet{ebort}) and deaths \textito{(i.e., a person in state $D$ or $B$)} in \textito{each} country $i$ at date $t$, denoted by NRC$(i,t)$ and NRD$(i,t)$, are given for dates starting on March, 23$^{\rm rd}$ 2014. \textitr{However, those report are not published daily. Thus, in order to complete the missing information we use a cubic hermite interpolation method assuming that \textitb{on November 1$^{\rm st}$,} 2013 all countries were free of \textitb{the} disease.} Thus, \textito{taking into account the characteristics of the EVD presented in Section \ref{ebola},} we estimate the amount of persons in \textito{states} $E$, $I$, $H$, $R$, $D$ and $B$ at $t=0$ \textito{as follows}: 
\begin{itemize}

\item \textito{Since the main duration in} state $H$ is 4.5 days, we compute \textito{$H(i,0)$} by considering the \textito{number of reported } cases \textito{that are} alive at \textito{time $t=0$} minus \textito{the number of reported cases that are alive \textito{4.5} days} before
 \textito{$$H(i,0) =\big((NRC(i,0)-NRD(i,0))-(NRC(i,-4.5)-NRD(i,-4.5))\big).$$}

\item \textito{Since the main duration in} state $D$ is 2 days, we compute $D(i,t)$ as $$D(i,0) =\big(NRD(i,0)-NRD(i,-2)\big)).$$

\item \textito{Since the main duration in states $E$, $I$ and $H$ is 11.4, 5 and 4.5 days, respectively,  we consider  
\textitm{$$E(i,0) =\dfrac{11.4}{4.5} TH(i,0) \hbox{ and } I(i,0) =\dfrac{5}{4.5} TH(i,0),$$}
where \textito{$TH(i,0)=NRC(i,0)-NRC(i,-4.5))$} is the \textito{total} number of persons who have \textito{ever} been in state $H$ \textito{any of the last 4.5 days}.}
%$$E(i,t) =\dfrac{\gamma_{H}(i)}{\gamma_{E}}\cdot TH(i,0) \hbox{ and } I(i,t) =\dfrac{\gamma_{H}(i)}{\gamma_{I}}\cdot TH(i,0),$$
%where $\gamma_{H}(i)=\big(\omega(i) \cdot \gamma_{HD}+(1-\omega(i)) \cdot \gamma_{HR}\big)$. the mean duration of a person in those states, the mean duration of a person in state $H$ in country $i$ denoted by $\gamma_{H}(i)$, $
\item The number of recovered persons $R(i,0)$ is given by
$$R(i,0) =\big(NRC(i,0)-NRD(i,0)-H(i,0) \big).$$
\item The number of buried cadaver $B(i,0)$ is 
$$B(i,0) =\big(NRD(i,0)-D(i,0) \big).$$
\item The number of susceptible persons $S(i,0)$ is 
$$S(i,0) =\big(NP(i,0)-E(i,0)-I(i,0)-H(i,0)-R(i,0)-D(i,0)-B(i,0)\big).$$
\end{itemize}
All these numbers are rounded to the nearest integer.

\subsubsection{\textitr{Rate \textitb{of movement} of people between countries} $\tau(i,j)$ \label{fluxe}}

\textitr{Regarding the dynamic of the current \textitr{2014-15} Ebola epidemic and the way the disease has spread from Guinea to other countries, it seems that the short-term visits (such as the visit to relatives) \textitb{are} the main causes of \textitb{EVD} diffusion between countries \citep{BMC}. %(http://www.biomedcentral.com/1741-7015/12/196)
 \textitb{However, such information} is quite difficult to obtain at a worldwide level, especially when considering African countries.} 

\textitr{Here, to approximate \textitb{the} pattern of those short term visits, we have considered the data regarding the 2005-2010 human migratory fluxes between countries obtained from \citet{migr} and freely available at the URL: 
\url{http://www.global-migration.info}
Following the analysis provided in \citet{ONS}, %(http://www.ons.gov.uk/ons/dcp171776_312055.pdf)
that compare short term immigration (a stay round 3 months) and the \textitb{census} in England, we assume that the pattern of the movement of people between countries of this migration database and the short term visits exhibit a similar general behavior (i.e., similar visited countries). However, it is expected that short term visits occur much more often than long-term immigration. Thus, considering both assumptions, we compute the percentage of persons in country $i$ \textitb{moving} to country $j$ per day as
\begin{equation*}
\tau(i,j)=c_{\tau} \tilde{\tau}(i,j)/(5 \cdot 365 \cdot NP(i,0)),
\end{equation*}
\textitb{where} $\tilde{\tau}(i,j)$ \textitb{represents} the number of persons \textitb{moving} from country $i$ to country $j$  from 2005 to 2010 and $c_{\tau} \geq 1$ is a parameter used to increase the value of $\tilde{\tau}(i,j)$.} 

\textitr{In addition to $c_{\tau}$, the parameter $\epsilon_{\rm fit}$, presented in Section \ref{BHT}, also play a role in the control of the movement of people as it does not allow a country $i$ to spread \textitb{EVD} to other countries until $E(i,t)\geq \epsilon_{\rm fit}$. In this work, we have estimated, by considering a \textitb{trial} and error process performed \textitb{in} the experiment presented in \textitb{Section \ref{valid}}, that \textitb{$\epsilon_{\rm fit}=2 \times 10^{-3}$} and $c_{\tau}=2.9$ produce a pattern of \textitb{EVD} spread between countries close to the one observed during the 2014-15 epidemic. This numerical result seems to indicate that the approach considered here regarding the movement of people is reasonable.}

\textitr{\textitb{Of course}, if data about short term visits are available they should replace the values of matrix $\tau$ used in this work.}

\subsubsection{EVD characteristics \label{EVDC}}
The following parameters are assumed to be well studied due to several data sets available about the current \textitr{2014-15} Ebola outbreak. Using data from Sections \ref{ebola} and \ref{cind}, we estimate the following parameters for our model: 
\begin{itemize}

\item $\mu_m(i)=1/\hbox{MLE}_i$ (day$^{-1}$)

\item \textitb{%$\omega(i,t)=0.25 \cdot \dfrac{\hbox{SAN}_i}{\max_i(\hbox{SAN}_i) }+ (0.728 \cdot m_I(i,t)) \cdot (1-\dfrac{\hbox{SAN}_i}{\max_i(\hbox{SAN}_i) })$
\textitf{$\omega(i,t)=\delta ((1-m_I(i,t)) \underline{\omega}+  m_I(i,t) \hbox{W}_i) +(1-\delta)\hbox{W}_i$}, \textitf{where we \textit{denote} by W$_i=  \dfrac{\hbox{SAN}_i}{\max_i(\hbox{SAN}_i)}    \underline{\omega}  +    (1-\dfrac{\hbox{SAN}_i}{\max_i(\hbox{SAN}_i)})  \overline{\omega}$} the disease fatality percentage of country $i$ when no control measure is applied; $\underline{\omega}\in[0,1]$ is the minimum disease fatality percentage; $\overline{\omega}\in[0,1]$ is the maximum disease fatality percentage; \textitm{and $\delta\in[0,1]$ denotes the proportion of the fatality percentage that can be reduced due to the application of control measures.} For the EVD, we consider \textitf{$\delta=0.53$,} $\underline{\omega}=0.25$ and $\overline{\omega}=0.728$. This formula and \textitf{the value of $\delta$} were determined empirically during the experiments presented in Section \ref{res} \textitb{so} that the cumulative number of deaths returned by our model and the real observations reported in \citet{ebo4} \textitm{were} similar. To determine the expression of $\omega(i,t)$, we have taken into account that,\textito{ for this particular epidemic,} the EVD fatality percentage oscillate between \textito{[25,72.8]\%}, depending on the quality of the sanitary service (see \citet{ebo9}). \textito{Moreover}, according to \citet{ebo4}, the maximum fatality rate \textitm{has decreased} with the application of the control measures. Thus, we have modelled this effect by multiplying the maximum fatality rate by $m_I(i,t)$.}

\item $\gamma_E=1/11.4$ (day$^{-1}$) and $\gamma_{D}=1/2$ (day$^{-1}$)

\item $C_{o}=$13 (days) is an upper bound estimation of the convalescence period computed as the maximum number of days that a persons stays hospitalized minus the duration of a person in the state $H$,

\item \textitb{$\gamma_I(i,t)=1/(d_I-g(i,t))$ (day$^{-1}$), $\gamma_{HR}(i,t)=1/(d_{HR}+g(i,t))$ (day$^{-1}$), $\gamma_{HD}(i,t)=1/(d_{HD}+g(i,t))$ (day$^{-1}$),  where $d_I$, $d_{HR}$ and $d_{HD}$ denote the mean duration in days of a person from state I to H, from state H to R and from state H to D, respectively, without the application of control measures; $g(i,t)=d_g \cdot (1-m_{I}(t,h))$ represents the decrease of the duration $d_I$ due to the application of control measures in country $i$ at time $t$; and $d_g$ is the maximum number of days that  $d_I$ can be decreased due to the control measures. Here, according to Section \ref{ebola}, we consider $d_I=5$, $d_{HR}=5$, $d_{HD}=4.2$ and $d_g=3$ (assuming that in cases of strong control measures, as in the United Kingdom, $d_I$ can be reduced by 3 days\textitb{, see} \cite{ebo4}).}

\item $\beta_I(i)$: There exists several works on the computation of the EVD effective contact rate $\beta_I(i)$ considering various SIR model (see \citet{ebo6,ebocongo}). However, the value of this rate depend on the epidemic characteristics (country, year, etc.). Furthermore, our model includes novel characteristics regarding those articles, as it includes movement between countries, hospitalized people and control measures. Thus, we have computed our own rates by using a regression method considering three particular sets of data associated with the evolution of the EVD epidemic in Guinea, Liberia and Sierra Leone (see \citet{ebo4,ebo7}).

\begin{itemize}
\item \textito{In Guinea, the country }of origin of the EVD epidemic, the index case was identified as a young boy who died on December 6$^{\rm th}$, 2013 and infected 3 persons of its family. \textitr{On July 14$^{\rm th}$, 2014, the estimated date for which the effects of the application of the control measures by national and international authorities start to slow down the \textitb{EVD} spread dynamic, a total cumulative number of 425 cases and 319 dead persons were reported (see \citet{ebo4}). After this date, the international help started to affect the initial EVD effective contact rate in Guinea, denoted by $\beta_I(\rm Guinea)$. }
Thus, we fit those data with the solution given by \textito{system} \eqref{SEIHR}. \textito{To this end,} \textito{system} \eqref{SEIHR} \textito{was} started at $t=0$ (corresponding to  \textito{December 6}$^{\rm th}$, 2013) with 3 persons \textito{in state} $I$ in Guinea, 1 person in state $D$ and all other persons being free of disease. The model \textito{was} run with $T_{\max}=220$ days (corresponding to  \textitr{July 14$^{\rm th}$, 2014} as the final date). In this \textitr{particular} simulation, we \textito{did} not consider control measures (i.e., for any $(i,j,t) \in \N \times \N \times \R$, $m_I(i,t)=m_H(i,t)=m_{\rm tr}(i,j,t)=1$). All other parameters \textito{were} set to the values introduced previously. Considering a particular value $\beta^I_{\rm Guinea} \in \R^{+}$, at the end of this simulation we \textito{computed} \textitr{the model error Err$(\beta^I_{\rm Guinea})=\sqrt{\hbox{$\sum_{t=1}^{T_{\max}}\big($cumul$_{\rm cases}$(Guiena,t)-NRC$($Guinea$,t)\big)^2$}}$}. We minimized Err$(\beta_I(\rm Guinea))$ by considering a dichotomy algorithm starting from $\beta_I(\rm Guinea)$=0.117 (day$^{-1}$) (\textitf{see} \citet{ebo6}) and found an optimal \textitr{value of $\beta_I(\rm Guinea)$=\textito{0.1944} (day$^{-1}$).}

\item In Sierra Leone, \textitr{8 cases and 2 deaths} were reported on \textitr{March 31$^{\rm th}$, 2014}. \textitr{The effects of control measures are estimated to start on September 9$^{\rm th}$, 2014,}when the cumulative number of reported cases was 2792 with 1501 deaths. We \textito{used} the same fitting technique as in the case of Guinea, with \textito{system} \textito{\eqref{SEIHR}, without control measures,} \textito{starting} at $t=0$ (corresponding to \textitr{March 31$^{\rm th}$, 2014}) with 3, 1, 1, 1, 5 and 1 persons in \textito{states} $E$, $I$, $H$, $D$, $R$ and $B$, respectively  (considering the estimation method presented in Section \ref{estim}) and all other persons being free of disease. \textitr{The system \textito{was} run with $T_{\max}=162$ days (i.e., final date \textitb{on}  \textito{September 9}$^{\rm th}$, 2014). We found
$\beta_I(\rm Sierra Leone)$\textito{=0.2605}} (day$^{-1}$). %This fitting process is presented in Figure \ref{?}.

\item \textitr{In Liberia,  16 cases and 5 deaths were reported on May 27$^{\rm th}$, 2014. On  \textitb{June} 30$^{\rm th}$, 2014, before the start of effects of the control measures, 302 cases and 139 deaths were observed.} System \textito{\eqref{SEIHR}, without control measures,} \textito{was} started at $t=0$ (i.e., corresponding to \textitr{May 27$^{\rm th}$, 2014}) \textitr{with 5, 2, 2, 1, 9 and 4 persons} in the states $E$, $I$, $H$, $D$, $R$ and $B$, respectively (see Section \ref{estim}). This system \textito{was} run during \textitr{$T_{\max}=34$  days}. Applying the same technique as for Guinea and Sierra Leone, we found $\beta_I(\rm Liberia)$\textito{=0.2649} (day$^{-1}$).

\end{itemize}

\textito{Taking into account} those three rates, since the rate of other countries (especially the non African ones) remains unknown (due to the lack of data), we have performed an empirical non linear regression to \textito{estimate $\beta_{I}(i)$. To this aim, $\beta_{I}(i)$ is assumed to be a} \textito{non-decreasing} function \textito{$\beta_{I}(r_{\beta} \cdot \hbox{DEN}_i/\hbox{GNI}_i)$,} where \textito{$r_{\beta}\in [0,+\infty)$} \textito{(km$^{2}$$\cdot$US\$$\cdot$persons$^{-2}$$\cdot $year$^{-1}$)} is a balance parameter which determines the importance of \textito{$\hbox{DEN}_i$} on the value of  $\beta_{I}$ in comparison to  \textito{$\hbox{GNI}_i$}. Indeed, the variable \textito{$r_{\beta}\cdot \hbox{DEN}_i/\hbox{GNI}_i$} is chosen because of the following reasons: 1) we assume that the \textito{higher the population of a country is}, the \textito{higher the probability of contagion is}\textito{ and the higher $\beta_I(i)$ is}; 2) the higher the economy level of \textito{a} country is, the higher \textito{its} \textito{education level is}, the lower the EVD risk habits of persons are (for instance, touching cadavers during funerals, see \citet{ecoebo,ebolaculture}) \textito{and the lower $\beta_I(i)$ is}. 
%We \textito{observe that} $\hbox{GNI}_{\rm Guinea}/\hbox{DEN}_{\rm Guinea}=5.49$ (km$^2$ $\cdot$ US\$ $\cdot$persons$^{-2}$ $\cdot$year$^{-1}$), $\hbox{GNI}_{\rm Sierra Leone}/\hbox{DEN}_{\rm Sierra Leone}=4.24$ (km$^2$ $\cdot$ US\$ $\cdot$persons$^{-2}$ $\cdot$year$^{-1}$) and $\hbox{GNI}_{\rm Liberia}/\hbox{DEN}_{\rm Liberia}=3.0236$ (km$^2$ $\cdot$ US\$ $\cdot$persons$^{-2}$ $\cdot$year$^{-1}$). 
\textito{In addition, we propose to use }a function of the form
\textito{$$\beta_{I}\Big(r_{\beta}\dfrac{\hbox{DEN}_i}{\hbox{GNI}_i}\Big)=a_{\beta}\arctan \Big(r_{\beta}\dfrac{\hbox{DEN}_i}{\hbox{GNI}_i}+b_{\beta}\Big)+c_{\beta},$$}
where $a_{\beta}$ (day $^{-1}$), $b_{\beta}$ \textito{(non-dimensional)}  and $c_{\beta}$ (day$^{-1}$) $\in \R$. We found, by considering the nonlinear regression method \emph{'nlinfit'} implemented in Matlab \textito{using} \textitr{\textito{the points }
\begin{itemize}
\item $(\hbox{DEN}_{\rm Guinea}/\hbox{GNI}_{\rm Guinea}, \beta_I(\rm Guinea))=(0.1820,0.1944)$, 
\item $( \hbox{DEN}_{\rm Sierra Leone}/\hbox{GNI}_{\rm Sierra Leone}, $ $\beta_I($Sierra Leone$))=(0.2357,0.2605)$,
\item $( \hbox{DEN}_{\rm Liberia}$ $/\hbox{GNI}_{\rm Liberia}, \beta_I($Liberia$))=(0.3307,0.2649)$,
%\item \textito{and the point (0,0) (i.e., we assume that in non populated areas the effective contact rate is null),}
\end{itemize}
that $a_{\beta}=105.3318   $, $b_{\beta}=-19.9227 $, $c_{\beta}=0.0328$ and $r_{\beta}=0.2156$.} 
\begin{remark}
If needed, $\beta_I$ can be also considered as time dependent \textito{(see \cite{Forgoston})}. For instance, as said in Section \ref{ebola}, it has been observed that Ebola Virus survives better outside the host for lower temperatures (see \citet{ebo8}). Thus, it could be interesting to introduce a slight dependence of $\beta_I(i)$ on the temperature of the country $i$. For instance, we could consider:
\textito{\begin{equation} \label{betafin}
\bar{\beta}_I(i,t)=\beta_{I}\Big(r_{\beta}\dfrac{\hbox{DEN}_i}{\hbox{GNI}_i}\Big)\Big(1-\alpha  \dfrac{\hbox{TMP}_i(t)-\hbox{TMP}_{\rm ref}}{\max_{(i,t)}|\hbox{TMP}_i(t)-\hbox{TMP}_{\rm ref}|}\Big),
\end{equation}}
where TMP$_{\rm ref}$ (ºC) is a reference \textito{temperature;} and $\alpha$ (\%) represent the maximum percent variation of the value $\beta_I$.
However, as no data  are available in literature to estimate a suitable value of $\alpha$, the effect of the temperature is neglected in our model.
\end{remark}
\textito{\item \textito{\textitr{$\beta_H(i)= \big(\beta_I(i)/25 \big)$} (day$^{-1}$), \textitm{since} the probability of being infected by contact with \textito{persons in state $H$} is \textitr{25} times lower than the probability of being infected by contact with \textito{persons in state $I$}, as explained in Section \ref{ebola}},}
\item \textito{$\beta_D(i)=\beta_I(i)$ (day$^{-1}$), \textitm{since} the probability of being infected by contact with \textito{persons in state $D$} is the same as the probability of being infected by contact with \textito{persons in state $I$} \textitm{(see Section \ref{ebola})}.}

\end{itemize}

\subsubsection{\textito{Control measures} \label{kappa}}
\textito{Here, we estimate the parameters used in Equation \ref{wcm}:}

\begin{itemize}
\item \textito{$\lambda(i)$: \textito{It is the first day $t$ such that $H(i,t) \geq 1$ for all countries except for Guinea, Liberia, Mali, Nigeria and Sierra Leone. For these countries, intensive control measures were not applied right after the apparition of the first person in state $H$, but some time later (as reported in \citet{ebo4}). Thus, in \textito{the} simulations presented in Section \ref{res}, \textito{we considered} for these countries a \textito{reported} delay between the detection of the first EVD case and the application of the control measures.}}

\item \textito{$\kappa_i$: In order to fit $\kappa_i$,} we \textitm{used} data from Guinea, Sierra Leone and Liberia. \textito{\textito{Moreover}, for all the numerical experiments considered in this work, we \textitm{considered} $m_{\rm tr}(i,j,t)=m_{I}(i,t)m_{I}(j,t)$.}

\begin{itemize}
\item In Guinea, \textitr{the effect of the control measures started on July 14$^{\rm th}$, 2014}. On \textitb{April 15$^{\rm th}$, 2015}, the number of reported cases in Guinea was 3568.
Again, we fit those data with \textito{system} \eqref{SEIHR} starting at $t=0$ (corresponding to \textito{December 6$^{\rm th}$,} 2013) with 3 persons \textito{in state} $I$ and 1 person in \textito{state} $D$ in Guinea and all other persons being free of the disease. The model \textito{was} run with $T_{\max}$=496 days (corresponding \textitb{to April 15$^{\rm th}$, 2015}). In this simulation, the control measures \textito{were} applied after \textitr{$t=220$ days (i.e., $\lambda$(Guinea)=220 days)}. Considering a particular value $\kappa_{\rm Guinea}$, we \textito{computed} the model error Err$(\kappa_{\rm Guinea})$, as defined previously. We minimized \textito{Err$(\kappa_{\rm Guinea})$} by considering a dichotomy algorithm starting from $\kappa_{\rm Guinea}$=0.001 and found an optimal value of \textitr{$\kappa_{\rm Guinea}$=0.00125} (day$^{-1}$).

\item In Sierra Leone, \textitr{the effects of control measures started} on \textitr{September 9$^{\rm th}$, 2014}. On
 \textitb{April 15$^{\rm th}$, 2015}, the number of reported cases in Sierra Leone was 12294. \textito{We used} the same fitting method as in Guinea and started \textito{system} \eqref{SEIHR} with the same conditions as those used for computing $\beta^I_{\rm Sierra Leone}$.  The system was started with \textitr{$T_{\max}$=389 days} and control measures \textito{were} applied at day \textitr{$\lambda$(Sierra Leone)=228} (i.e., corresponding to \textitr{September 9$^{\rm th}$, 2014)}. We found \textitr{$\kappa_{\rm Sierra Leone}$=0.00227}.

\item In Liberia, the effects of control measures started on \textitr{June 30$^{\rm th}$, 2014}. \textitb{On
April 15$^{\rm th}$, 2015,} the number of reported cases was 10241. We used the same fitting method as the one used for Guinea and started \textito{system} \eqref{SEIHR} with the same conditions as those used for computing $\beta^I_{\rm Liberia}$. This system was run with \textitr{$T_{\max}$=333 days} and control measures were applied at day \textitr{$\lambda_{\rm Liberia}$=34}. We found \textitr{$\kappa_{\rm Liberia}$=0.00270}.
\end{itemize}

\textito{Taking into account those three values, we perform a regression method, similar to the one introduced in Section \ref{EVDC}, 
estimating $\kappa_{i}$. To this end, $\kappa_{i}$ is assumed to be a \textito{non-decreasing} function $\bar{\kappa}(r_{\kappa} \cdot \hbox{SAN}_i/\hbox{DEN}_i)$, where \textito{$r_{\kappa}\in [0,+\infty)$ (persons$^{2}$$\cdot $year$\cdot$ km$^{-2}$$\cdot$US\$$^{-1}$)} is a balance parameter which determines the importance of $\hbox{SAN}_i$ on the value of  $\beta_{I}$ in comparison to  $\hbox{DEN}_i$. Indeed, the variable $r_{\beta}\cdot  \hbox{SAN}_i/\hbox{DEN}_i$ is chosen because of the following reasons: 1) the higher the sanitary expenses are, \textito{the more efficient} the control measures are \textito{and the higher \textito{$\kappa_{i}$} is}; 2) \textito{the higher the value of $\hbox{DEN}_i$ is}, the  harder to respect the control measures \textito{is and the lower \textito{$\kappa_{i}$} is}. %We note that $\hbox{SAN}_{\rm Guinea}/\hbox{DEN}_{\rm Guinea}=5.70\times 10^{-8}$ (km$^2$ $\cdot$ US\$ $\cdot$persons$^{-2}$ $\cdot$year$^{-1}$), $\hbox{SAN}_{\rm Liberia}/\hbox{DEN}_{\rm Liberia}=3.42\times 10^{-7}$ (km$^2$ $\cdot$ US\$ $\cdot$persons$^{-2}$ $\cdot$year$^{-1}$) and $\hbox{SAN}_{\rm Sierra Leone}/\hbox{DEN}_{\rm Sierra Leone}=1.86\times 10^{-7}$ .
} 

Again, we propose to use 
$$\bar{\kappa}\Big(r_{\kappa}\dfrac{\hbox{SAN}_i}{\hbox{DEN}_i}\Big)=a_{\kappa}\arctan\Big(r_{\kappa}\dfrac{\hbox{SAN}_i}{\hbox{DEN}_i}+b_{\kappa}\Big)+c_{\kappa}\,$$ 
where $a_{\kappa}$ (day $^{-1}$), $b_{\kappa}$ \textito{(non-dimensional)}  and $c_{\kappa}$ (day$^{-1}$) $\in \R$. We found, by considering the nonlinear regression method \emph{'nlinfit'} implemented in \textito{Matlab } \textito{\textito{using the points}
\textitr{\begin{itemize}
\item $( \hbox{SAN}_{\rm Guinea}/\hbox{DEN}_{\rm Guinea}, \bar{\kappa}(\rm Guinea))$=(0.6697,0.00125), 
\item $(\hbox{SAN}_{\rm Sierra Leone}/\hbox{DEN}_{\rm Sierra Leone},$ $\bar{\kappa}( $Sierra Leone))=(1.1344,0.00227), 
\item $(\hbox{SAN}_{\rm Liberia}/$ $\hbox{DEN}_{\rm Liberia}, \bar{\kappa}(\rm Liberia))$=(1.4686,0.00270),  
%\item and the point (0,0) (i.e., we assume that no sanitary expense implies no control measures), 
\end{itemize}}
that \textitr{$a_{\kappa}=0.0132$, $b_{\kappa}=-0.3222 $, $c_{\kappa}=-0.1523$ and $r_{\kappa}= 0.0475$}.} 

However, regarding the evolution of the control measures during the current EVD epidemic, since the beginning of August the international community have sent important sanitary and financial help to affected countries to help them to eradicate the EVD outbreaks. Thus, we assume that all countries affected by EVD will have a control measure coefficient $\kappa_{i}$ at least as efficient as \textitr{$\kappa_{\rm Liberia}$}. Thus, we consider 
\textito{$$\kappa_{i}=\max\Big(\bar{\kappa}\Big(r_{\kappa}\dfrac{\hbox{SAN}_i}{\hbox{DEN}_i}\Big),\kappa_{\rm Liberia} \Big).$$ }
\end{itemize}

%%%%%%%%%%%%%%%%%%%%%%%%%%%%%%%% NUMERICAL EXPERIMENT

\subsection{Numerical experiments \label{res}} 
We consider the parameters presented in Section \ref{parameter} and carry out several numerical experiments in order to estimate some relevant values of the \textitr{2014-15} Ebola outbreak. First in Section \ref{RID},  \textito{we study the initial risk} of introduction and diffusion of \textito{EVD for each country,} \textito{taking into account only some of the inputs of the model}. Next, in Section \ref{valid}, we validate our \textito{model} by comparing the outputs of two numerical experiments with real data. Finally, in Section \ref{temporal}, we predict the possible EVD evolution \textitb{starting from recent data and up to the end of the epidemic}. 

\subsubsection{Initial Risk of EVD Introduction and Diffusion \label{RID}}

Here, we consider $t=0$ corresponding to \textitr{April 24$^{\rm th}$, 2015}, the date of the last EVD situation report available in \citet{ebo4} when those experiments were performed. \textitm{For the countries \textitf{affected by EVD} on \textitr{April 24$^{\rm th}$, 2015}, \textito{we \textitf{used} the initial conditions presented in Table \ref{dat2}\textitf{,} obtained by using the methodology presented in Section \ref{estim} and the values of the parameters given in Sections \ref{kappa} and \ref{fluxe}}.}

\textitor{\begin{remark} \label{Ldel}
We note that the last EVD case in Liberia was reported on March 22$^{\rm th}$, 2015  and this country was declared free of disease on May 9$^{\rm th}$, 2015 (see \cite{ebo4}). However, in the reports presented in \cite{ebo4}, due to the delay in receiving the results from laboratories to confirm past EDV cases, the table reporting the cumulative number of cases and deaths is still \textitf{given} up to April 24$^{\rm th}$, 2015. Thus, in this work, for all experiments starting from April 24$^{\rm th}$, 2015, Liberia was \textitf{considered free} of disease.   
\end{remark}}

\begin{table}
\begin{center}
\caption{\textito{Initial conditions for the countries affected by EVD on \textitr{April 24$^{\rm th}$, 2015}.}  \label{dat2}}
\textitr{\begin{tabular}{l|rrrrrrr}
\hline
\textbf{Country}& $E(i,0)$ & $I(i,0)$ &  $H(i,0)$ & $R(i,0)$ & $D(i,0)$ &  $B(i,0)$\\
\hline
Sierra Leone&182&80&70&8397&6&3889\\
%Liberia			&202&89&79&5576&20&4596\\
Guinea       &25&11&10&1201&6&2368\\
\hline
\end{tabular}}
\end{center}
\end{table}
%set:\textito{ $E$(Guinea,0)$=145$,  $ \lambda({\rm  Guinea})=-256$;  $E$(Liberia,0)$=222$, $ \lambda({\rm Liberia})=-123$; $E$(Sierra Leone,0)$= 1007$, $ \lambda($Sierra Leone$)=-131$; $E$(USA,0)$=0$,  $ \lambda({\rm USA})=-66$;  $E$(Mali,0)$=0$, $ \lambda({\rm  Mali})=-43$; $E$(Nigeria,0)$=0$ $ \lambda({\rm Nigeria})=-134$;  $E$(Senegal,0)$=0$ $ \lambda({\rm Senegal})=-98$; $E$(Spain,0)$=0$ $ \lambda({\rm  Spain})=-60$}.

\textito{In} Table \ref{RDI}, \textito{we present} the value of RS \textito{for the} countries with a strictly positive value of RS on \textitr{April 24$^{\rm th}$, 2015}. \textitr{\textito{We observe that Guinea \textitf{has} a risk value around \textito{$3.0\times 10^{-4}$}} persons sent to other countries per day} (i.e., a probability of \textito{0.03\%} to contaminate another country per day). \textitr{The risk value of Sierra Leone is lower. Those values are quite low and seem to indicate that a new spread of Ebola outside those \textito{two} countries have a low probability to occur.} 
\begin{table}
\begin{center}
\caption{Risk of EVD spread to other countries (RS) on \textitr{April 24$^{\rm th}$, 2015 for countries affected by EVD.}  \label{RDI}}
\textitr{\begin{tabular}{lrrr}
\hline
Country&RS\\
\hline
Guinea               &3.0$\times 10^{-4}$\\
%Liberia              &2.8$\times 10^{-4}$\\
Sierra Leone         &1.5$\times 10^{-4}$\\
\hline
\end{tabular}}
\end{center}
\end{table} 

In Table \ref{RII}, we report the 20 countries with the highest value of RI. We \textito{observe} that Liberia and Sierra Leone are the first countries in this list, which is logical since those countries receive an important \textitr{number of people} from Guinea. This result is consistent with the \textitr{2014-15} Ebola situation, with the disease starting from Guinea and then spreading to Liberia and Sierra Leone. We also \textito{observe} that the USA, \textitf{the United Kingdom,} Spain, Nigeria and Senegal are in this top 20 (all \textitb{were} affected by EVD cases). Indeed, they receive an important \textitr{amount of persons} from \textito{Guinea and Sierra Leone}. The \textito{USA, France and \textitf{the United Kingdom} \textito{had (on \textitr{April 24$^{\rm th}$, 2015})} a probability \textito{of receiving} an infected person per day of \textito{0.0007\%}}, this value is more than twice higher than the risk of other countries in the list (except Liberia and Sierra Leone). \textitr{Again, those values are low and tend to show that the current \textitb{EVD} epidemic should not spread anymore outside Guinea, Liberia and Sierra Leone.}

\textitr{Another interesting feature of models such as Be-CoDiS are their ability to estimate the magnitude of an epidemic in a country free of disease. Thus, for each country  reported on Table \ref{RII} which is currently free of \textitb{EVD} (i.e.,\textito{ we exclude Sierra Leone}), we have run the model starting from one infected individual in the considered country at day $t=0$ (corresponding to \textitr{April 24$^{\rm th}$, 2015}) and all other countries free of disease. The model stops when the epidemic ends (i.e., the approximated time for which the numbers of persons in states $E$, $I$, $H$ and $D$ are all lower than 1). In Table \ref{RII}, we show for those countries the final value of cumul$_{\rm cases}$ and cumul$_{\rm deaths}$ described in Section \ref{cout}. We observe that, in case of spread of \textitb{EVD} to those countries, no major outbreak (i.e., more than 10 reported cases) should be noticed, except for \textito{Gambia, Nigeria and Guinea Bissau with 94, 36 and 20 reported} cases, respectively.}

\begin{table}
\begin{center}
\caption{Risk of EVD introduction (RI) on \textitr{April 24$^{\rm th}$, 2015} for each country. Only the Top 20 is reported. \textitr{Furthermore, for countries which are actually free of \textitb{EVD} disease (i.e., all countries \textito{expect Sierra Leone}), we report the cumulative number of cases (\textbf{Cases}) and deaths (\textbf{Deaths}) predicted by Be-CoDiS in cases of \textitb{EVD} introduction.}  \label{RII}}
\textitr{\begin{tabular}{lrrr}
\hline
\textbf{Country}&\textbf{RI}&\textbf{Cases} &\textbf{Deaths}\\
\hline
\textito{Liberia }       &\textito{3.5$\times 10^{-4}$}&\textito{4}&\textito{1}\\
Sierra Leone            &3.5$\times 10^{-5}$&-&-\\
United States of America&7.9$\times 10^{-6}$&2&0\\
France                  &6.8$\times 10^{-6}$&2&0\\
United Kingdom          &6.6$\times 10^{-6}$&2&0\\
Spain                   &3.7$\times 10^{-6}$&2&0\\
Senegal                 &2.3$\times 10^{-6}$&6&3\\
Australia               &2.2$\times 10^{-6}$&2&0\\
Canada                  &2.1$\times 10^{-6}$&2&0\\
Netherlands             &1.8$\times 10^{-6}$&3&0\\
Italy                   &1.7$\times 10^{-6}$&2&0\\
Gambia                  &1.6$\times 10^{-6}$&94&56\\
Germany                 &1.4$\times 10^{-6}$&2&0\\
Portugal                &1.3$\times 10^{-6}$&2&0\\
Mauritania              &1.1$\times 10^{-6}$&2&0\\
South Africa            &7.6$\times 10^{-7}$&2&0\\
Nigeria                 &6.8$\times 10^{-7}$&36&11\\
Belgium                 &6.6$\times 10^{-7}$&2&0\\
Sweden                  &5.8$\times 10^{-7}$&2&0\\
Guinea Bissau           &5.6$\times 10^{-7}$&20&12\\
\hline
\end{tabular}}
\end{center}
\end{table}

Such a study is interesting as it reveals the countries with the most immediate risk of EVD introduction or spread. Effort for controlling the \textitf{movement of people} entering or leaving those regions \textito{could} be prioritized in order to reduce the spread of EVD \textitb{worldwide}.

%A world map showing the distributions of RI for each country is included in Figure \ref{DRI}. We see that Western Europe, Western Africa, North America and Australia \textito{were} the areas \textito{with the highest risk of EVD introduction}.
%\begin{figure}
%\begin{center}
%\includegraphics[width=11.89cm]{RIc.eps}
%\caption{Risk of EVD Introduction (RI) of each country, corresponding to data on December 7$^{\rm th}$, 2014. Darker zones correspond to higher risk values.\label{DRI}}
%\end{center}
%\end{figure}

\subsubsection{Validation of the model \label{valid}}

\textitr{We are now interested in validating \textitm{the Be-CoDiS model by} considering the \textitm{results obtained for the 2014-15 EVD epidemic}. In particular, we first want to check its ability \textitm{for} generating \textitm{forecasts for} long time intervals (i.e., from the beginning to the end of the epidemic). Then, after recalibrating some model parameters by considering \textitb{only some} recent epidemic data, we study the behavior of our model in the case of prediction for short term intervals (\textitm{less} than 2 months).}

\paragraph{Validation for long time intervals:} We considered a simulation starting from the known index cases of the current EVD epidemic on December 6$^{\rm th}$, 2013 and let the epidemic run until its ending (i.e., the approximated time for which the numbers of persons in states $E$, $I$, $H$ and $D$ are all lower than 1). To \textito{this aim}, \textitb{system} \eqref{SEIHRc} \textito{was} started at $t=0$ (corresponding to December 6$^{\rm th}$, 2013) with 3 persons \textito{in state} $I$ and 1 person \textito{in state $D$} in Guinea and all other persons being free of the disease. \textito{All the parameters \textito{were} set to the values introduced previously in Section \ref{parameter}. \textitb{In particular,} taking into account Section \ref{kappa}, we set the delay between the first day $t$ such that $H(i,t)\geq 1$ and the first day of application of control measures for \textitr{Guinea, Sierra Leone and Liberia to 220, 220 and 180 days}, respectively.}

\begin{figure}
\begin{center}
\includegraphics[width=11.89cm]{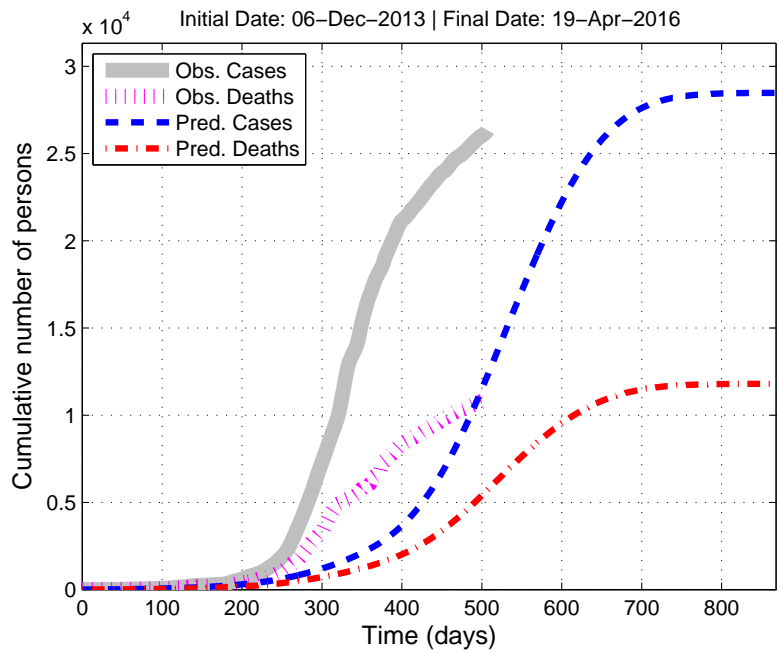}
\caption{\textitb{Evolution of the cumulative numbers of total EVD cases (dashed line) and deaths (dash-dotted line) predicted by the Be-CoDiS model from December 6$^{\rm th}$, 2013 to \textitb{April 19$^{\rm th}$, 2016}. We also show the cumulative amount of total cases (continuous line) and deaths (dotted line) reported in \citet{ebo4}.}\label{ltpred}}
\end{center}
\end{figure}

\begin{figure}
\begin{center}
\includegraphics[height=7cm]{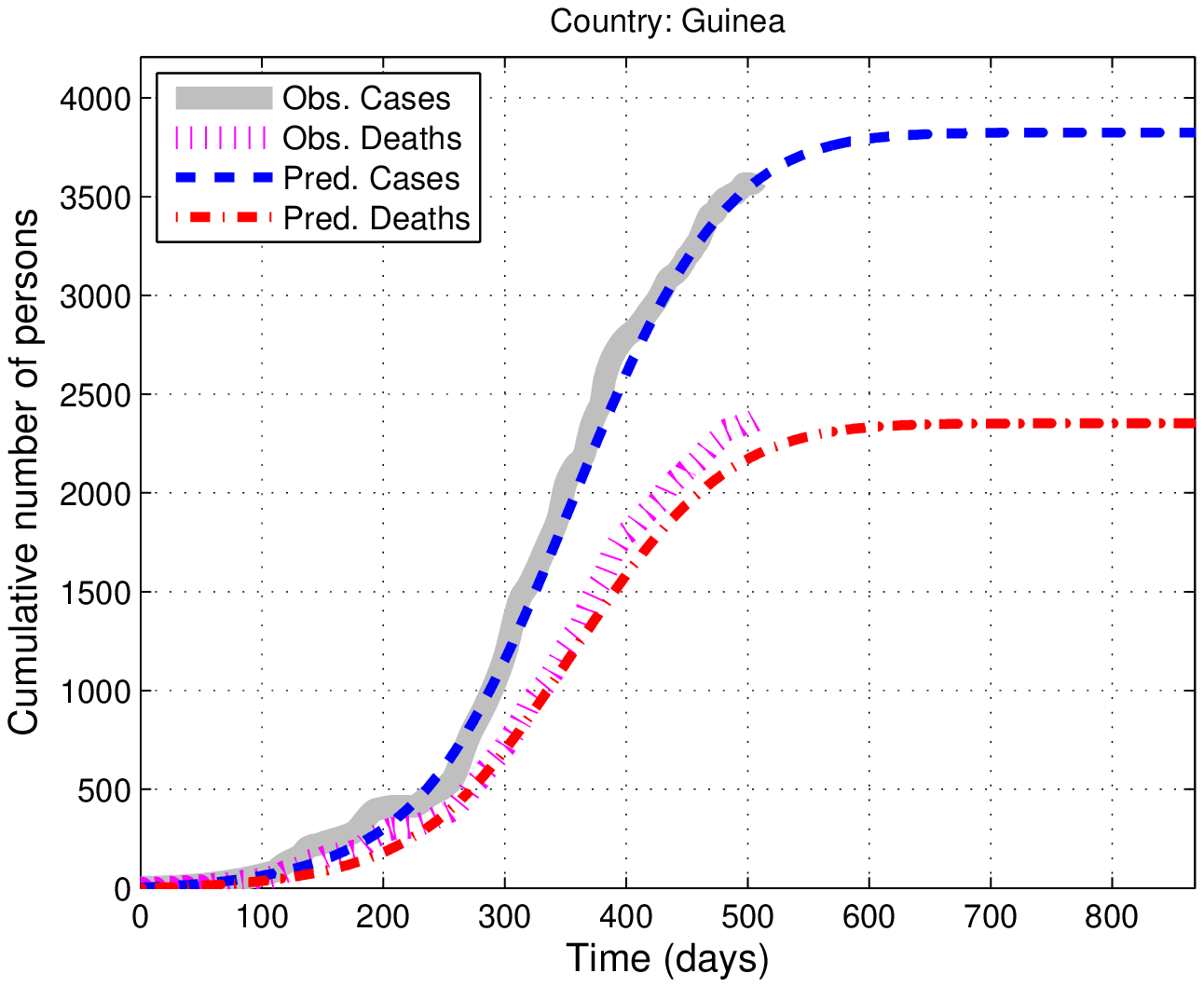}\\
\includegraphics[height=7cm]{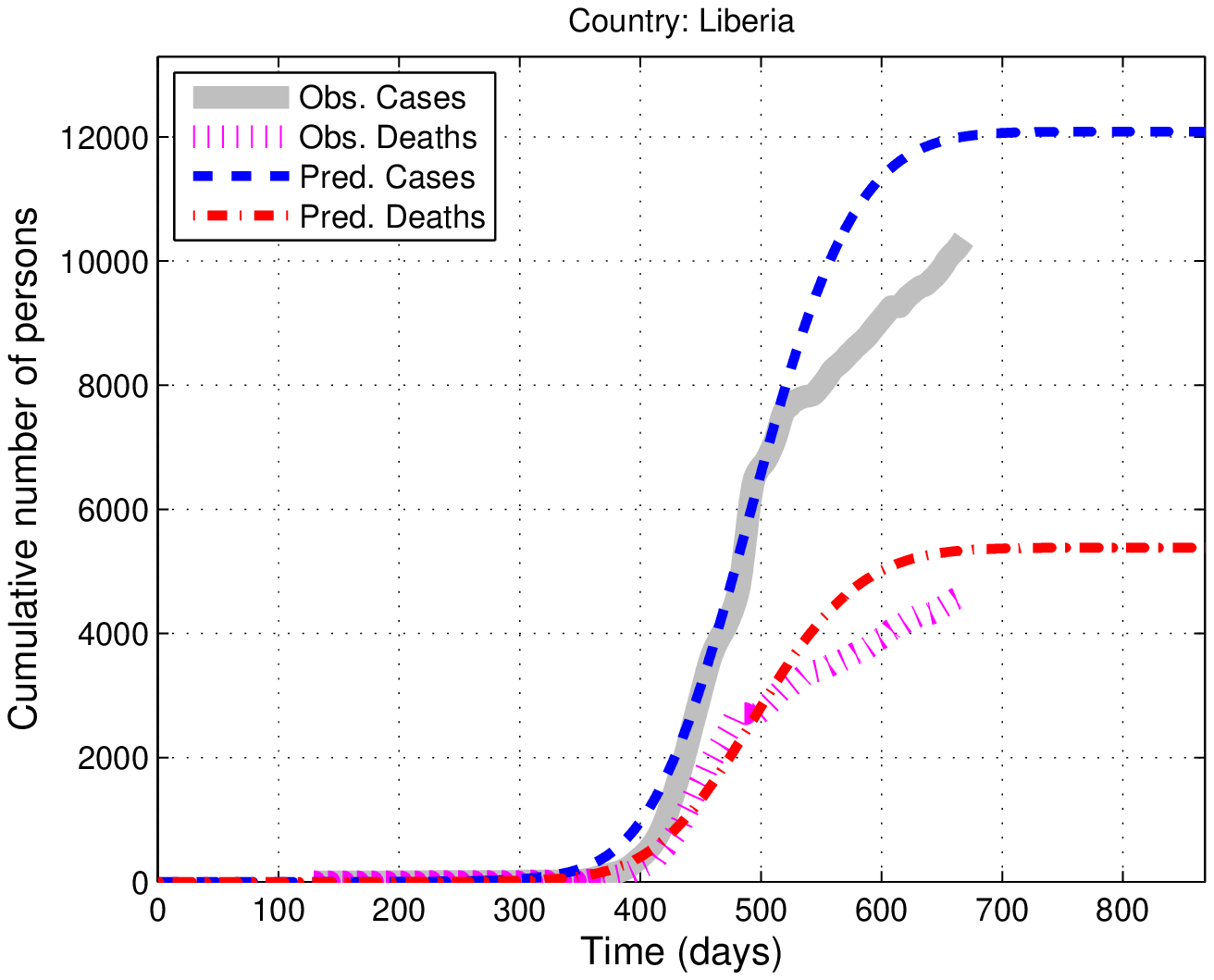}\\
\includegraphics[height=7cm]{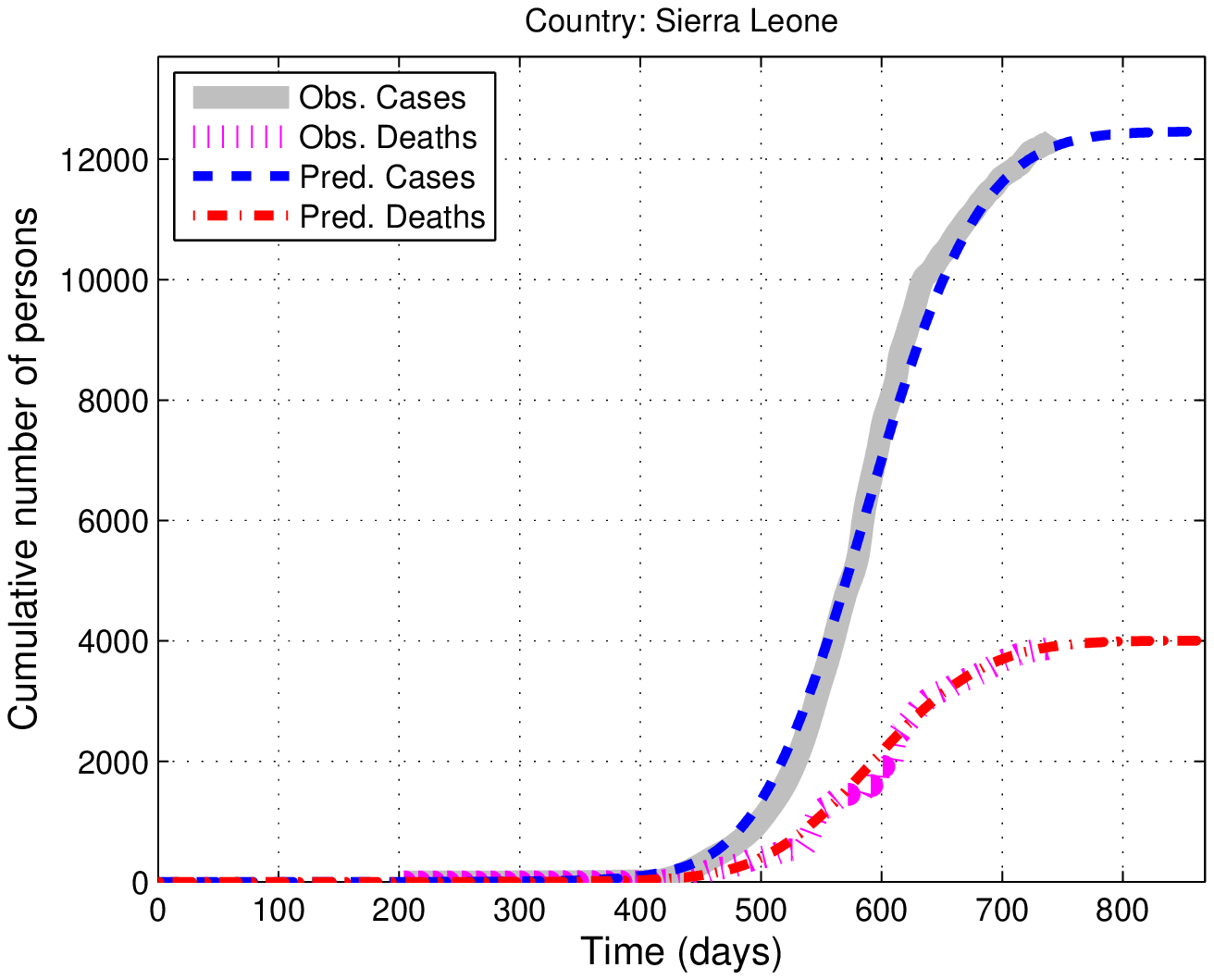}
\caption{\textitb{Evolution of the cumulative numbers of EVD cases (dashed line) and deaths (dash-dotted line) predicted by the Be-CoDiS model from December 6$^{\rm th}$, 2013 to \textitb{April 19$^{\rm th}$, 2016} for (\textbf{Top}) Guinea, (\textbf{Middle}) Liberia and (\textbf{Bottom}) Sierra Leone. We also show the cumulative amount of cases (continuous line) and deaths (dotted line) reported in \citet{ebo4}. The dates of observed data have been translated in order to fit them with the initial date of infection in each country predicted by the model.}\label{ltpredma}}
\end{center}
\end{figure}

\textitr{The evolution of the cumulative numbers of total cases and deaths predicted by the model is presented in Figure \ref{ltpred}. We also show \textito{in this} figure the cumulative numbers of cases and deaths observed by the authorities during the epidemic (see \citet{ebo4}) for several dates and interpolated by cubic Hermite polynomials to obtain a continuous representation of the data. In addition, we depict in Figure \ref{ltpredma} the evolution of the cumulative  numbers of cases and deaths of the three most affected countries (i.e, Guinea, Sierra Leone and Liberia) predicted by the model and their corresponding observed data. We note that in this figure, due to the delay of the model in spreading the EVD between countries, which is explained below, we have translated the dates of the observed data in order to fit them with the initial date of infection in each country predicted by the model. We \textito{point out} that the real observations are also based on estimations done by the authorities and are periodically corrected (for instance, they count suspected cases of EVD and remove them if the serological EVD tests are negative). This explain the oscillation in the cumulative numbers of cases and deaths.}

In addition, in Table \ref{Tltpred}, we report \textitr{the R$_0$ and CR$_0$ values of the model and affected countries,} the date of the first infection (i.e., the first time for which the cumulative number of infected cases is greater than 1), the final cumulative numbers of cases and the final cumulative numbers of deaths, and the maximum number of persons hospitalized at the same time in each country (i.e., MNH) for countries affected by EVD predicted by \textito{Be-CoDiS. \textitb{Moreover},} \textito{we also show the data reported in \citet{ebo4}  for those countries for \textitr{April 24$^{\rm th}$, 2015}}. 

\begin{table}
\begin{center}
\caption{Validation for long time intervals: R$_0$ and CR$_0$ values (\textbf{R}$_0$), \textitf{date} of the first reported case (\textbf{Date}), cumulative numbers of cases (\textbf{C.}) and deaths (\textbf{D.}) and maximum number of hospitalized persons \textitf{MNH} (\textitf{\textbf{H.}}) for countries affected by EVD predicted by Be-CoDiS (\textbf{BC}) and in the real epidemic (\textbf{Real}). Real observed data reported in \citet{ebo4} on \textitr{April 24$^{\rm th}$, 2015} are also shown.  \label{Tltpred}}
\textitr{\begin{tabular}{l|rrrrr|rrr}
\hline
 & \multicolumn{5}{c}{\textbf{BC}}  & \multicolumn{3}{|c}{\textbf{Real}}\\
\hline
\textbf{Country}& \textbf{R}$_0$&\textbf{Date} & \textbf{C.} &  \textbf{D.} &\textitf{\textbf{ H.}} & \textbf{Date} & \textbf{C.} &  \textbf{D.} \\
\hline
Total           &2.1&-&28475& 11797& 2231 & -& 26302 &  10899\\
\hline
Sierra Leone    &2.4&27-07-14& 12461&4007 &  1033&26-05-14& 12371 & 3899\\
Liberia         &2.3&04-06-14& 12087& 5383&	991&31-03-14& 10322 & 4608\\
Guinea          &1.8&06-12-13&  3825&2353 &  175&06-12-13& 3584& 2377\\
Nigeria         &2.5&22-05-15& 21&7       &  2&20-07-14 &20  &8\\
Senegal         &1.8&18-02-16& 1&0        &  1&29-08-14 &1&0\\
USA             &1.5&09-10-15& 1&0        &  1&30-09-14 &4 &1\\
UK             &1.6&11-11-15& 1&0         &  1&30-09-14 &4 &1\\
\hline
Gambia          &2.4&14-01-15& 78&47	     &  6& - &0 &0\\
\hline
Mali            &-&-&0&0&0&23-10-14&8&6 \\
Spain           &-&-&0&0&0&06-10-14&1&0 \\
\hline
\end{tabular}}
\end{center}
\end{table}

From Table \ref{Tltpred}, we \textito{observe} \textitr{that our model predicts the infection of Sierra Leone, Liberia, Nigeria, Senegal, the USA and \textitf{the United Kingdom}.} Moreover, the epidemic magnitude (i.e., final number of cumulative cases and deaths)  in those countries, in Guinea and summing all affected countries is similar to the one observed on  \textitr{April 24$^{\rm th}$, 2015}.  In addition, on the one side, our model also \textitm{forecasts} the infection of Gambia with \textito{relatively} low epidemics, which has not occurred before this work was done. On the other side, \textito{starting just with data from December 6$^{\rm th}$, 2013},  our model fails to predict infection \textitr{in Mali and Spain.} However, \textito{when this work was done, }in those four countries the EVD epidemic \textito{seemed} to be sporadic and limited. \textitr{Regarding the dates of first infection of each affected country, we can observe that the spread of \textitb{EVD} to other countries present a delay when comparing them to the real situation. This delay can be explained by a low estimation of the movement of people. However, regarding the data used during this work no better estimation of the \textitf{movement of people} has been obtained. 
Focusing on the MNH values, the model \textitm{predicts} that at least 2231 beds in hospital should be required to treat all affected peoples at the same time. 
The R$_0$ \textitm{value}, defined in \textitb{Section \ref{cout}}, \textitm{is also given} in this table and is around 2.1. This value \textitb{coincides} with \textitm{those} reported in \cite{ebocomp} \textitm{and \cite {ebo5}} computed before \textitb{October, 2014} (R$_0=2.0$). This \textitm{seems} to show that the general dynamic of the epidemic predicted by our model \textitm{is similar to} the one estimated \textitf{in} those \textitm{other works}.} 

\textitr{\textit{In} Figure \ref{ltpred}, we see that the model \textitf{estimates} a global magnitude of the epidemic similar to the observed data. \textitm{However, the global evolution of the simulated epidemic is slower than \textitf{the evolution} reported in \citet{ebo4}.} This can be explained by the delay mentioned previously in the spread of \textitb{EVD} between countries. However, we observe \textitm{in} Figure \ref{ltpredma} that, once the country is infected, the evolution of the epidemic in Guinea, Sierra Leone and Liberia have a similar behaviour to the observed one. When considering only the most recent observations, we \textitm{see} that, as expected, the difference between observed and simulated values increase with time. This is particular visible for Liberia\textitf{,} where an important change in the epidemic dynamic has been observed \textitm{in} December 2014 due to the quick positive effects of control measures\textitm{,} whereas for Guinea and Sierra Leone the \textitb{EVD} reached isolated areas and \textitm{is} still highly active up to February 2015 \textitm{(see \cite{ebo4})}. This \textitm{seems} to indicate that some of the model parameters should be recalibrated in order to produce more accurate forecast \textitm{when} considering only the most recent data.}

\textitr{From those results, and taking into account the difficulty of epidemiological models for predicting long time intervals (see \citet{massad}), Be-CoDiS seems to generate reasonable behaviour of the evolution of the epidemic when considering a long time interval, even when using just the initial index cases as initial data. In particular, it seems that the model may help to detect, from the start of an epidemic to its end, the countries with more probabilities to develop important outbreaks and to simulate the relevant values of affected cases. To our knowledge no similar experiment was done by other models proposed in the literature.  However, we also see some limitations of the model related to the speed of the spread of the \textitb{EVD} between countries and the difference between final estimated values and real observations. }


\paragraph{Validation for short time intervals:} \textitr{As said previously, we are now interesting in recalibrating some model coefficients by taking into account only the most recent data. To this aim, as since February, 2015 the epidemic dynamic has slightly decreased, we have considered data from \textitm{February 1$^{\rm st}$,} 2015 up to \textitb{April 24$^{\rm th}$, 2015} \textitb{(the last date available when those experiments were run)}. We choose to recalibrate parameters $\beta_I$ and $\kappa$, which are the ones that determine the dynamic of the epidemic due to human interactions and control. Other parameters are associated with the natural process of \textitb{EVD} and should not vary in time. Then, we optimize the values of $\beta$ and $\kappa$ for Guinea, Liberia and Sierra Leone by minimizing the function \textitb{TE$(\{\beta_I(i),\kappa(i)\}_{i\in\Sigma})=\hbox{$\sum_j \sum_{t=0}^{T_{\max}}\big($cumul$_{\rm cases}$(j,t)-NRC$($j$,t)\big)^2$}$, where $\Sigma=\{$Guinea, Liberia, Sierra Leone\},} with the hybrid genetic \textitb{algorithm (and its parameters)} presented in \citet{pof1}. To compute TE$(\{\beta_I(i),\kappa(i)\}_{i\in\Sigma})$, we run \textito{system} \eqref{SEIHRc} with \textito{$T_{\max}$=82 days} starting from February, 1$^{\rm st}$ 2015, with  the model parameters introduced in Section \ref{parameter} (except the values of $\beta_I$ and $\kappa$ which are given by the optimization method and the value of $\underline{\omega}$=0.32 to better fit the last observed death rates) and the initial conditions presented in Table \ref{dat1} obtained by using the methodology presented in Section \ref{estim}.}

\begin{table}
\begin{center}
\caption{\textito{Initial conditions for the countries affected by EVD on \textitr{February, 1$^{\rm st}$ 2015}.}  \label{dat1}}
\textitr{\begin{tabular}{l|rrrrrrr}
\hline
\textbf{Country}& $E(i,0)$ & $I(i,0)$ &  $H(i,0)$ & $R(i,0)$ & $D(i,0)$ &  $B(i,0)$& $\lambda(i)$\\
\hline
Sierra Leone&382&168&147&7317&21&3255&-508\\
Liberia			&233&102&91&4908&18&3728&-610\\
Guinea      &108&47&43&988&11&1933&-622\\
\hline
\end{tabular}}
\end{center}
\end{table}

\textitr{By considering this optimization process, we found that the $\beta$ values of Guinea, Liberia and Sierra Leone are 0.3117, 0.9616 and 0.2815, respectively and \textitb{the} $\kappa$ values are 0.0017, 0.0106 and 0.0042, respectively.}

\begin{figure}
\begin{center}
\includegraphics[width=11.89cm]{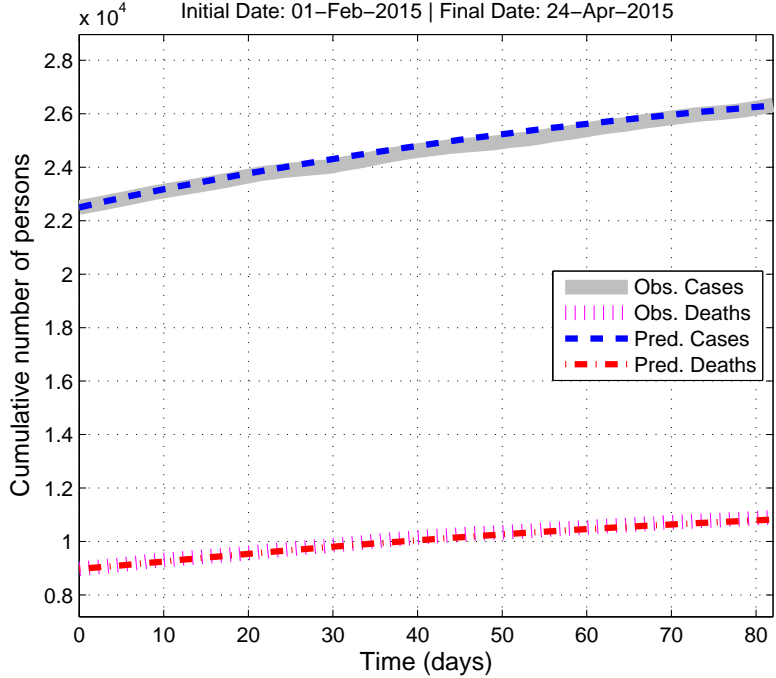}
\caption{\textitb{Evolution of the cumulative numbers of total EVD cases (dashed line) and deaths (dash-dotted line) predicted by the Be-CoDiS model from February 1$^{\rm st}$, 2015 to \textitb{April 24$^{\rm th}$, 2015}. We also show the cumulative amount of total cases (continuous line) and deaths (dotted line) reported in \citet{ebo4}.}\label{stpred}}
\end{center}
\end{figure}

\begin{figure}
\begin{center}
\includegraphics[height=7cm]{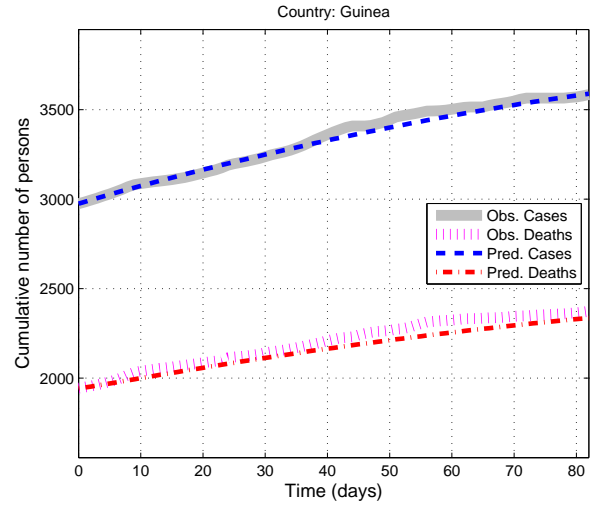}\\
\includegraphics[height=7cm]{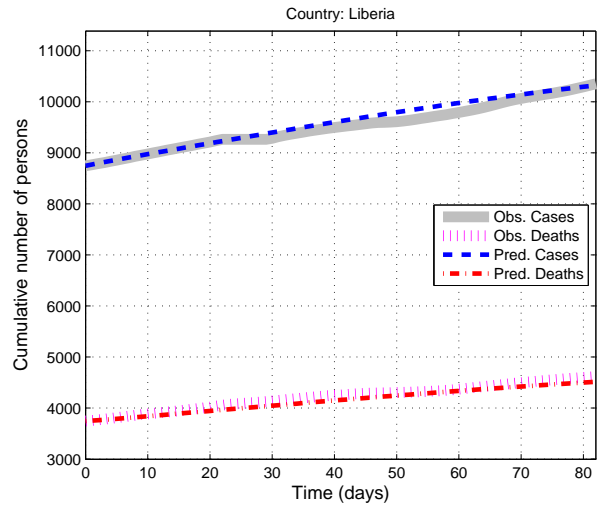}\\
\includegraphics[height=7cm]{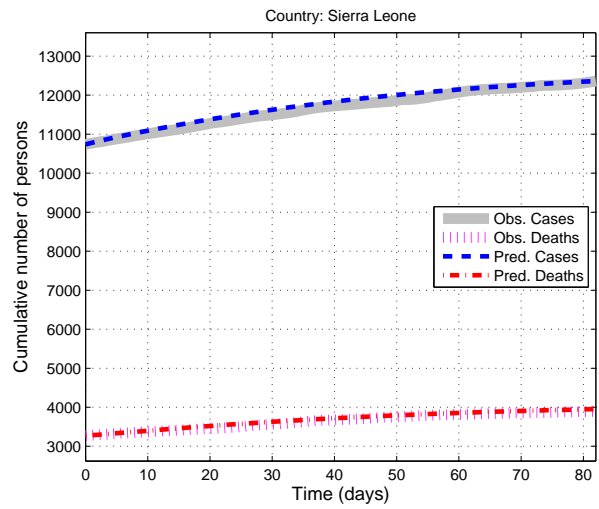}
\caption{\textitb{Evolution of the cumulative numbers of EVD cases (dashed line) and deaths (dash-dotted line) predicted by the Be-CoDiS model from February 1$^{\rm st}$, 2015 to \textitb{April 24$^{\rm th}$, 2015} for (\textbf{Top}) Guinea, (\textbf{Middle}) Liberia and (\textbf{Bottom}) Sierra Leone. We also show the cumulative amount of cases (continuous line) and deaths (dotted line) reported in \citet{ebo4}. }\label{stpredma}}
\end{center}
\end{figure}

\textitr{The evolution of the cumulative numbers of total cases and deaths predicted by the model and the \textitb{asssociated} real observations are presented in Figure \ref{stpred}. In addition, in Figure \ref{stpredma} the same values are reported for Guinea, Liberia and Sierra Leone. 
In Table \ref{Tstpred}, we summarize \textitm{the simulated} and observed cumulative numbers of cases and deaths for each affected country on \textitb{April 14$^{\rm th}$, 2015}\textitm{, the R$_0$ value of the model and the CR$_0$ values of each affected country}. In addition, we \textito{computed} and show in this table the percentage error of the model defined as ED(i)=mean value$_{t \in [ 1,T_{\max} ]} \big($cumul$_{\rm cases}(i,t)$-NRC$(i,t)\big)$/NRC$(i,t)$.}


\begin{table}
\begin{center}
\textitr{\caption{Validation for short time intervals: R$_0$ and CR$_0$ values (\textbf{R}$_0$), Cumulative numbers of cases predicted by Be-CoDiS (\textbf{BC}), observed cumulative cases reported in \citet{ebo4} (\textbf{RC}), percentage error on cumulative cases done by the model (\textbf{EC}), cumulative numbers of deaths predicted by Be-CoDiS (\textbf{BD}), observed cumulative deaths reported in \citet{ebo4} (\textbf{RD}), percentage error on cumulative deaths done by the model (\textbf{ED}) on \textitb{April 24$^{\rm th}$, 2015} for countries affected by EVD at those dates. We also report the cumulative number of deaths.  \label{Tstpred}}
\textito{\begin{tabular}{l|r|rrr|rrr}
\hline
\textbf{Country}& \textbf{R}$_0$ &\textbf{BC} &  \textbf{RC} & \textbf{EC (\%)}  & \textbf{BD} &  \textbf{RD} & \textbf{ED (\%)}  \\
\hline
Total &2.1&26307 & 26302& <1&  10817 & 10899& 1 \\
\hline
Guinea&1.9& 3589  & 3571&<1 &2336 &2370&2\\
Liberia&2.6& 10322 & 10322& 0&4513 & 4608&2  \\
Sierra Leone&1.8& 12361 & 12362& <1&3953& 3895&1 \\
\hline
\end{tabular}}}
\end{center}
\end{table}

\textitr{We observe \textitm{in} both\textitm{,} \textito{Figure \ref{stpred} and Table \ref{Tstpred}}, that the model fits quite well the observed data reported in \cite{ebo4}. \textitf{We note that for Liberia, the model reproduces the delayed laboratory confirmed EVD cases reported in \cite{ebo4}, since, as said in Remark \ref{Ldel}, the ultimate suspected EVD case in this country was observed on March 22$^{\rm th}$, 2015.}
Regarding the relative error done in predicting the number of cases and deaths, we observe that the model predicts \textitb{the final} number of cumulative cases with an error lower than 1\% and the final number of cumulative deaths with an error lower than 2\%. We also note that the considered R$_0$ value of the model remains unchanged with this new set of parameters values. This seem to indicate that our model predicts\textitm{, in a reasonable way,} the dynamic of the most recent observed data and, thus, we can use this new set of \textitb{parameters} to forecast the future evolution of the outbreak.}

 %Finally, focusing on the R$_0$ and CR$_0$ values computed during this experiment with the methodology presented in Section \ref{?}, we see that they are still elevated (increasing or decreasing, depending on the country, with respect to the long time interval validation experiment) but due to control measures the epidemic is controlled with an increase of cases in the considered 84 days (1832 new cases) reduced when regarding, for instance .

\subsubsection{Forecast for the \textitr{\textitr{2014-15}} EVD epidemic starting with data \textito{for} \textitr{April 24$^{\rm th}$, 2015}  \label{temporal}}

\textitr{In order to propose a forecast of the possible evolution of the \textitr{\textitr{2014-15}} EVD epidemic, \textitb{system} \eqref{SEIHRc} \textito{was} started with \textito{the parameters presented in Section \ref{parameter}} \textito{and the initial conditions reported in Table \ref{dat2} obtained by using the methodology presented in Section \ref{estim}.} The system \textito{was} run until the end of the epidemic (i.e., \textito{the time} for which the numbers of persons in states $E$, $I$, $H$ \textito{and $D$} are all lower than 1) which is estimated to be around \textitf{September 18$^{\rm th}$, 2015}. \textitf{We recall that, as specified in Remark \ref{Ldel}, Liberia is considered as free of disease. For this country, we consider the initial cumulative numbers of EVD cases and deaths corresponding the values reported on May 5$^{\rm th}$, 2015 (the last date available before the end of this work)}.} 

\begin{figure}
\begin{center}
\includegraphics[width=11.89cm]{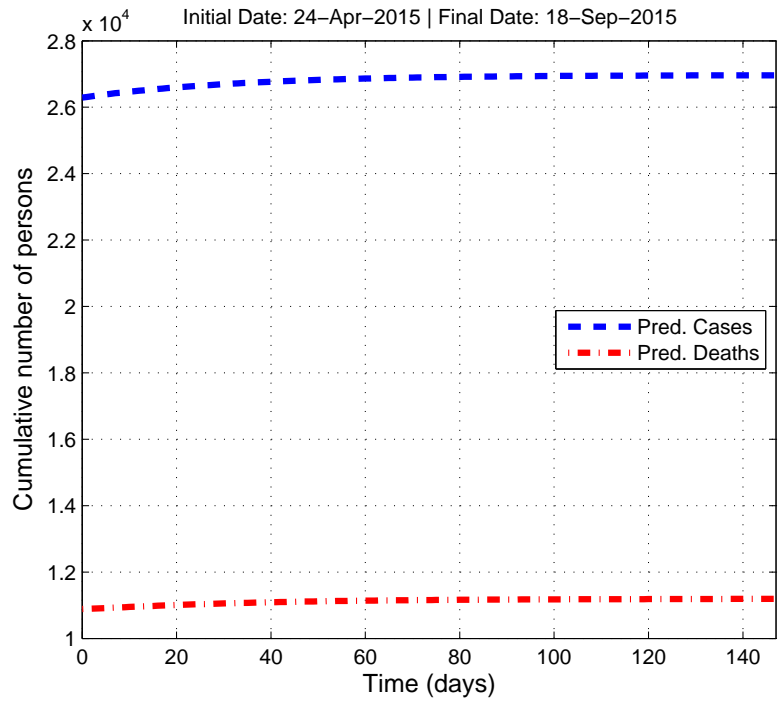}
\caption{Evolution of the cumulative numbers of total EVD cases (\textitb{dashed line}) and deaths (\textitb{dash-dotted line}) predicted by the Be-CoDiS model from \textitb{April 24$^{\rm th}$, 2015} to \textitf{September 18$^{\rm th}$, 2015}. \label{stevo}}
\end{center}
\end{figure}

\begin{figure}
\begin{center}
\includegraphics[height=8cm]{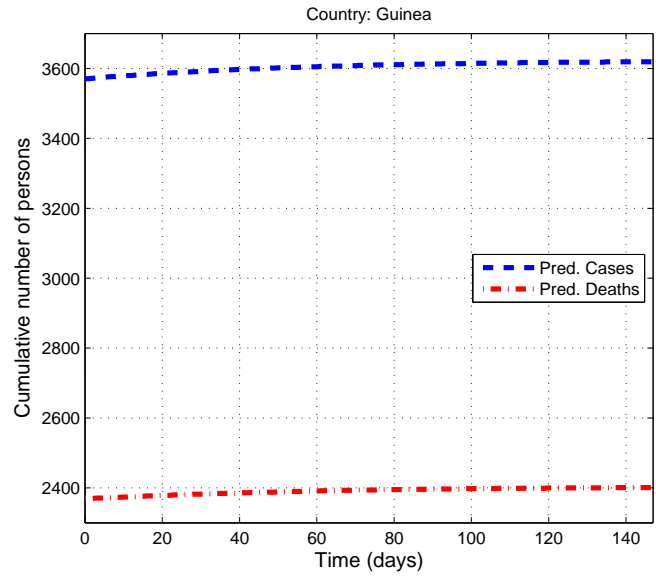}\\
\includegraphics[height=8cm]{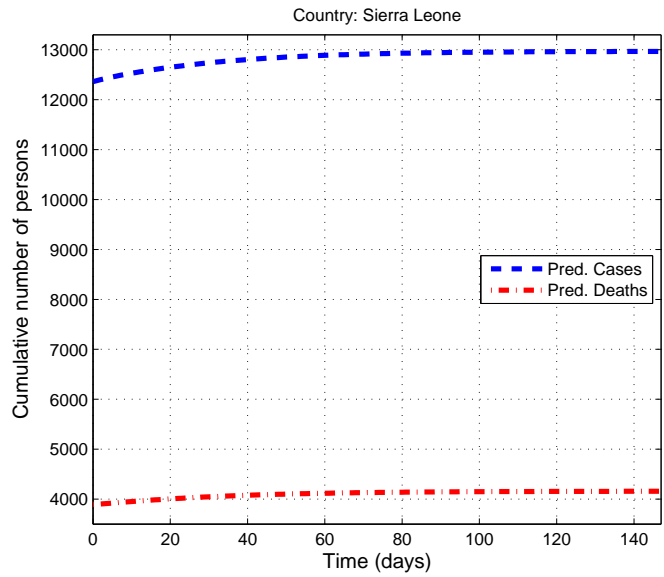}
\caption{\textito{Evolution of the cumulative numbers of EVD cases (dashed line) and deaths (dash-dotted line) predicted by the Be-CoDiS model from \textitb{April 24$^{\rm th}$, 2015} to \textitf{September 18$^{\rm th}$, 2015} for (\textbf{Top}) Guinea and (\textbf{Bottom}) Sierra Leone.  }\label{stevoam}}
\end{center}
\end{figure}

\begin{figure}
\begin{center}
\includegraphics[width=11.cm]{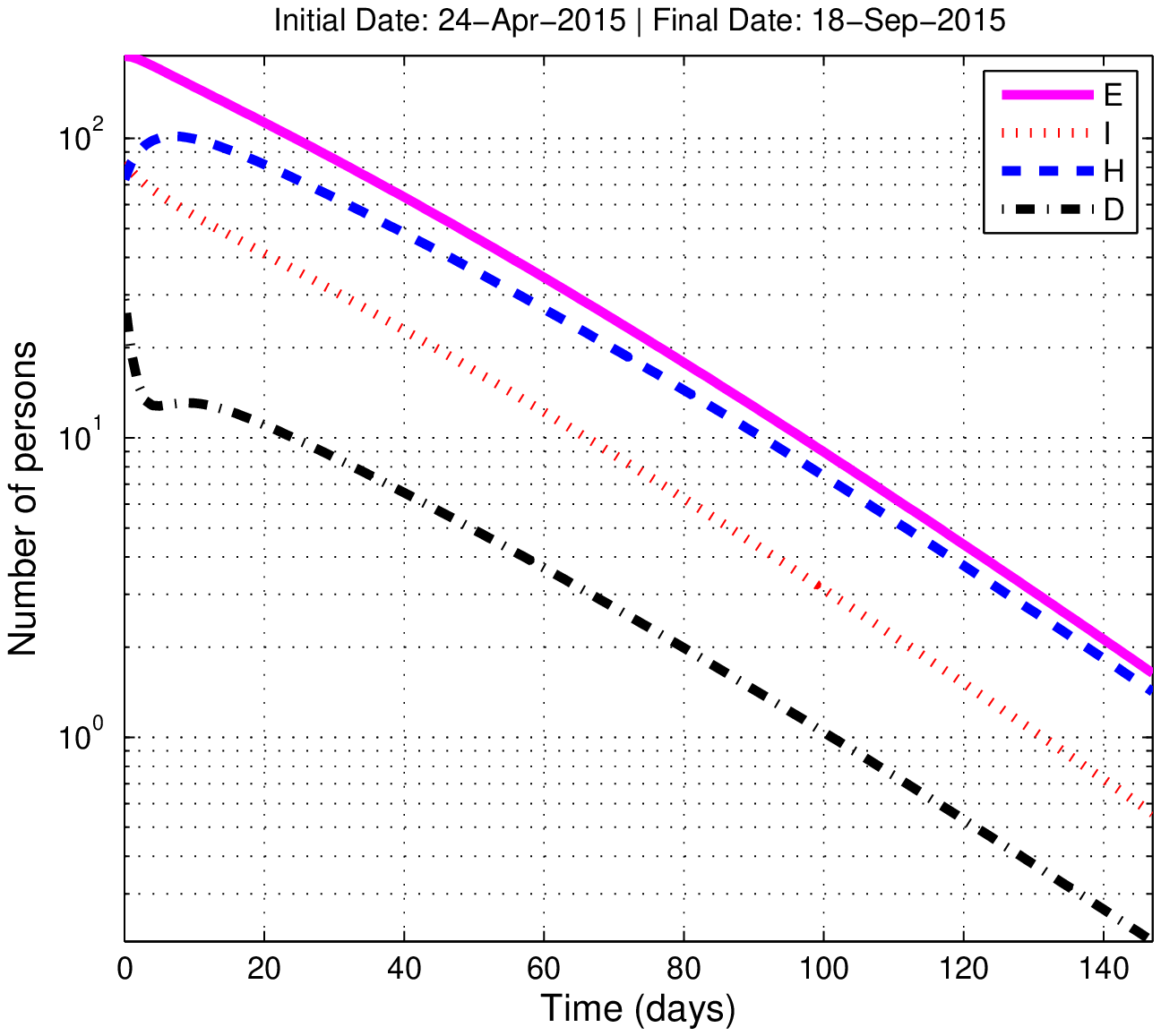}
\caption{Evolution of the total numbers of persons in states $E$ (dashed line), $I$ (slash-dashed line), $H$ (continuous line) and $D$ (slashed line) predicted by the Be-CoDiS model from \textitb{April 24$^{\rm th}$, 2015} to \textitf{September 18$^{\rm th}$, 2015}. \textitb{The Y-axis scale is logarithmic.} \label{stpevo}}
\end{center}
\end{figure}

\textitr{The evolution of the cumulative numbers of total cases and deaths predicted by the model is presented in Figure \ref{stevo}. Those cumulative numbers are also presented for \textitf{Guinea and Sierra Leone} in Figure \ref{stevoam}. The evolution of the number of persons \textito{in states} $E$, $I$, $H$ and $D$ is shown in Figure \ref{stpevo}. 
The final \textito{cumulative} numbers of cases and deaths determined by Be-CoDiS are \textitf{27243} and \textitf{11261}, respectively. 
The epidemic becomes stabilized with a slope that decreases progressively. This tendency seems to be confirmed by the last reported data (see \citet{ebo4}). This is also \textito{clear} in Figure \ref{stpevo}, where \textito{it can be seen that} the \textitf{total} number of person in states $E$, $I$, $H$ and $D$ decreases \textito{significantly}.}

\textitm{The countries} with a number of persons in \textito{states} $E$, $I$, $H$ or $D$ greater than 1 at \textito{least at one moment} from  \textitr{April 24$^{\rm th}$, 2015} to  \textito{November 23$^{\rm th}$, 2015} and their  characteristics \textitr{are} reported in Table \ref{sttab}. 
\textitr{We observe on this table that the risk of spreading EVD between countries during the whole period is low, with values \textitb{lower} than \textitf{0.5\%} (100$\times$TRS). Regarding the magnitude of the epidemic, the model \textito{predicts that} the epidemic \textitm{shows} a \textitb{decreasing} dynamic with a total of \textitf{659} new cases until the end of the \textitf{epidemic}. This decline can be observed in all affected countries.  \textitf{We note that, due to the movement of people from Guinea and Sierra Leone to Liberia, a new outbreak of 19 cases occurs in Liberia.}  
Regarding the values MNH of each country, we see that a maximum number of \textitf{217} beds in hospitals dedicated to treat EVD cases should be required to \textito{take care of} the affected persons. This value is much lower than number of the beds currently available \textitb{(\textitm{around} 1100 beds were available at January, 7$^{\rm th}$, 2015, \textitm{according to} \citet{ebo4})}. All those values tend \textitb{to be} consistent with the last observations reported in \cite{ebo4}, in which a clear \textitb{deceleration} of the epidemic is observed.}

\begin{table}
\begin{center}
\textitr{\caption{Forecast for the \textitr{\textitr{2014-15}} EVD epidemic starting with data \textito{for} \textitr{April 24$^{\rm th}$, 2015}: cumulative numbers of EVD cases (\textbf{Cases}),  cumulative numbers of deaths (\textbf{Deaths}), maximum number of persons hospitalized at the same time (\textbf{MNH}) and TRS values for countries affected by EVD predicted by Be-CoDiS on \textitf{September 18$^{\rm th}$, 2015}. We also report the cumulative \textito{number} of EVD cases (\textbf{I.C.}) and the cumulative \textito{number} of deaths (\textbf{I.D.}) \textito{reported} on \textitb{April 24$^{\rm th}$, 2015}\textitf{, except \textit{for} Liberia \textit{whose} values correspond to May 5$^{\rm th}$, 2015}. \label{sttab}}
\textito{\begin{tabular}{l|rrrr|rr}
\hline
\textbf{Country}& \textbf{Cases} &  \textbf{Deaths} & \textbf{MNH} & \textbf{TRS}& \textbf{I.C.} &  \textbf{I.D.} \\
\hline
Total                   &\textitf{27243}&\textitf{11261}& 217& - &\textitf{26584}&\textitf{10971}\\
Sierra Leone            &12966& 4157& 207&4.$\times 10^{-3}$&12362&3895\\
Liberia                &\textitf{10623}& \textitf{4778}& 1  &2.$\times 10^{-4}$&\textitf{10604}&\textitf{4769}\\
Guinea                  &3619 & 2401&  9 &5.$\times 10^{-3}$&3571 &2370\\
\hline
\end{tabular}}} 
\end{center}
\end{table}

\begin{figure}
\begin{center}
\includegraphics[width=11.cm]{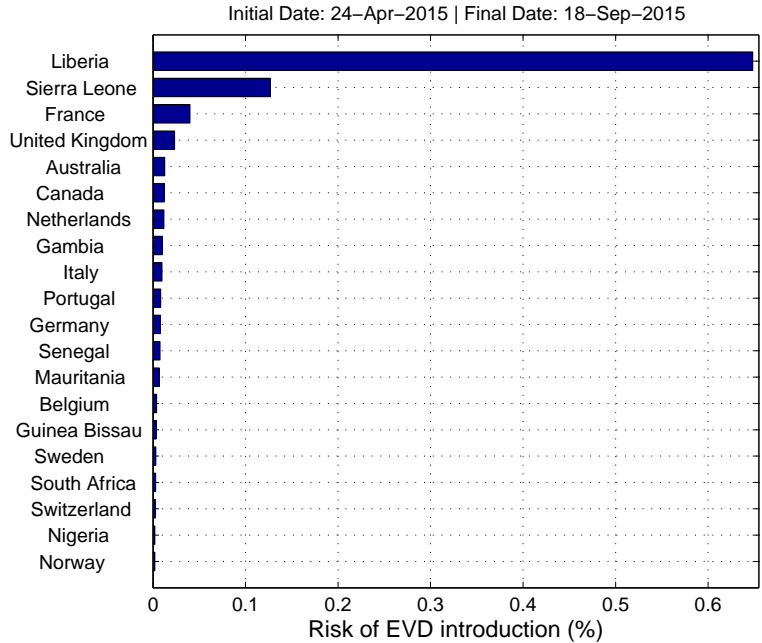}
\caption{Countries with the 20 highest probabilities (\%) of introduction \textito{of EVD, predicted by the Be-CoDiS model} from \textitb{April 24$^{\rm th}$, 2015} to \textitb{November 23$^{\rm th}$, 2015}. \label{sttri}}
\end{center}
\end{figure}

\textitb{Finally in Figure \ref{sttri}, we present a bar representation of the countries with the 20 highest values of TRI. More precisely, we report $100 \times$ TRI, as this value represents the probability of introduction of at least 1 EVD case due to \textitb{movement of people} (\%). \textitf{We see that Liberia is the country that \textit{has} the highest \textitf{probability} (around 0.7\%) to receive an infected person \textito{until the end of the epidemic}.} Considering other countries\textit{,} this probability is \textitb{lower} than \textitf{0.15\%, Sierra Leone,} \textitf{the United Kingdom} and France being the next countries  with the highest probabilities. This classification seems to be consistent with the results found in \citet{ebocomp}, where France and \textitf{the United Kingdom} are \textit{some} of the countries with the highest probabilities to receive an infected person. However, in \textitm{that} article, the probabilities of infection were around 75\% and was done in September 2014 when the EVD \textitb{epidemic} was very active.}

\subsection{Model sensitivity analysis \label{san}} 
The goal of this section is to provide a quick analysis of the variation  of the Be-CoDiS outputs regarding perturbations on the input data. To do so the model \textito{was run 100 times}\textitr{, with the values used for the forecasting experiment,} considering random uniform perturbations of amplitude $[-\alpha,+\alpha]$\% on all the parameters, with $\alpha=$1\%, 5\%, 10 \% and 20\%. We \textito{considered} the short time \textito{interval case studied in Section} \ref{temporal}. \textito{We computed the} mean percentage variations, considering all countries, of TRS, TRI, MNH, \textito{cumul$_{\rm cases}$ and cumul$_{\rm deaths}$,} regarding their respective non perturbed \textito{value}. For each value of $\alpha$, we report the average minimum, maximum, median and mean value considering all those variables. Results are reported in Table \ref{Tsan}.

\begin{table}
\begin{center}
\caption{Results of the sensibility analysis presented in Section \ref{san}. We report the mean, median, minimum and maximum percentage variation of the considered outputs regarding their non perturbed value. \label{Tsan}}
\textitr{\begin{tabular}{lrrrrr}
\hline
$\alpha$ & Mean& Mininmum &Maximum \\
\hline
1\%& 0.9 & 0.07 & 3.9 \\
5\%& 3.8 & 0.5 & 15.1 \\
10\%& 7.3 & 0.7 & 28.3\\
\hline
\end{tabular}}
\end{center}
\end{table}

From this table, we observe that perturbations of $\alpha$\% in the inputs generates mean output variations \textito{lower than} $\alpha$\%. Therefore, it seems that there is a linear relationship between input and output perturbations. The maximum observed mean perturbation is \textito{26.1\%} and is obtained for $\alpha$=10\%. This indicates that important variations in results can be obtained if input data are not good enough. A more extensive sensitivity analysis should be performed in order to identify the more influential model parameters and, thus, give recommendation about the result precision to possible users. 

\section{Conclusions \label{con}} 

In this work, we have presented \textitm{the} formulation of a new \textitm{deterministic} spatial-temporal epidemiological model, called \textito{Be-CoDiS,} based on the combination of \textitm{an Individual-Based} model (modelling the interaction between countries, considered as individual) for between country spread with \textitm{a compartmental} model, based on ordinary differential equations, for within-country spread.  The main characteristics of this model are the combination of the effects of the \textitr{movement of people} between countries and control measures and the use of dynamic model coefficients \textitr{adapted to each country}. The model has been validated considering the current \textitr{2014-15} EVD epidemic that strikes several countries around \textitr{the world}. 

Considering \textitr{the} validation experiments, the model reproduces in a reasonable way the real epidemic evolution. 
Starting from the index cases \textitb{in Guinea at December 2013}, we see that Be-CoDiS simulates a between country spread close to the real one. Furthermore, the magnitude of the simulated epidemic is similar to the observations reported in \cite{ebo4}. \textitr{However, we also observe a delay in the dates of between country spread and discrepancies considering the most recent data.  
Those results seem to indicate the validity of our approach but \textitm{also that} the model parameters should be \textitm{recalibrated} to better fits the last observed data. \textitb{Then, after recalibrating some model parameters with recent data, the model fits quite well the current epidemic dynamic.}}

\textitr{Regarding the forecast done by the model starting from  April 24$^{\rm th}$, 2015,} \textito{the epidemic should} disappear within \textitf{five} months and \textitr{the \textitm{decrease} of the number of cases in \textitf{Guinea and Sierra Leone} should \textitm{continue}.}   According to the model, the \textitr{final} magnitude of the epidemic could reach a total of \textitf{27243} \textito{cases, with \textitf{11261} deaths. \textitf{Due to movement of \textit{people}, Liberia, which \textit{has} no reported EVD \textit{cases} since March 22$^{\rm th}$, 2015, may have a new sporadic outbreak of 19 cases.} \textitf{Additionally, for each country except Liberia, the current risks of EVD introduction \textit{and} spread due to the movement of people are low and no EVD \textit{cases} should be reported outside Guinea, Sierra Leone and Liberia.} Moreover, the model estimates that \textitb{the} maximum number of beds required in hospital \textitr{should be} \textitf{217}.}

Finally, we have performed a brief sensitivity analysis of our model that seems to indicate a linear relation between perturbation in the inputs and outputs. However, in some cases, high \textitb{variations} can be obtained.

In this work, we have also highlighted the current limitation of our approach: simplified assumptions in the mathematical model, lack of precision in some data \textitr{(e.g., data used for the movement of people)} and the use of empirical assumptions \textitr{(e.g., the model uses regression formulas to obtain the parameters of some countries)}. 
Those parts should be improved in the future. 

\textitr{The next steps should be to perform a} more extensive sensitivity analysis \textitr{of this model in order} to identify the \textitb{key} parameters that have a strong impact on the model outputs. \textitr{Additionally,} the model can be used to study the economical impact of the \textitr{2014-15} Ebola epidemic and to solve associated optimization resource problems (for instance, controlling the epidemic considering a constrained economical \textito{budget}).

\textitr{A free Matlab version of the model presented here, including all the inputs related to the considered \textitb{EVD} epidemic, can be downloaded at the following URL:} 
\url{http://www.mat.ucm.es/momat/software.htm}

\section*{Acknowledgments} 
This work was carried out thanks to the financial support of  the Spanish ``Ministry of Economy and Competitiveness'' under project MTM2011-22658; the ``Junta de Andaluc\'ia'' and the European Regional Development Fund through project P12-TIC301;\textito{ and the research group MOMAT (Ref. 910480) supported by ``Banco de Santander'' and ``Universidad Complutense de Madrid''}.

% BibTeX users please use one of
%\bibliographystyle{spbasic}      % basic style, author-year citations

%\bibliographystyle{spbasic}      % basic style, author-year citations
%\bibliographystyle{spmpsci}      % mathematics and physical sciences
%\bibliographystyle{spphys}       % APS-like style for physics
%\bibliographystyle{plainnat}
%\bibliographystyle{spbasic} 
%\bibliography{becost}{}

%\end{thebibliography}

\end{document}